\documentclass[manuscript, screen]{acmart}
\usepackage{mdframed}
\usepackage{booktabs}
\usepackage{multirow}
\usepackage{listings}
\usepackage{stfloats}
\usepackage{enumitem} 
\usepackage{float}
\usepackage{subcaption}
\AtBeginDocument{%
  }

\setcopyright{acmlicensed}
\copyrightyear{2018}
\acmYear{2018}
\acmDOI{XXXXXXX.XXXXXXX}

\acmJournal{JACM}
\acmVolume{37}
\acmNumber{4}
\acmArticle{111}
\acmMonth{8}

\begin{document}

\title{CASPER — Change-Aware Slice Prioritization for Efficient Regression Testing of LLM-based systems}

\author{Biruk Asmare Muse}
\correspondingauthor
\email{bmuse@uottawa.ca}
\orcid{0000-0001-8861-9526}
\affiliation{%
  \institution{School of EECS, University of Ottawa}
  \city{Ottawa}
  \state{Ontario}
  \country{Canada}
}

\author{Lionel Briand}
\orcid{0000-0002-1393-1010}
\correspondingauthor
\email{lbriand@uottawa.ca}
\affiliation{%
  \institution{School of EECS, University of Ottawa}
  \city{Ottawa}
  \state{Ontario}
  \country{Canada}
}

\author{Yiwei Lu}
\orcid{0000-0001-7872-3186}
\correspondingauthor
\email{yiwei.lu@uottawa.ca}
\affiliation{%
  \institution{School of EECS, University of Ottawa}
  \city{Ottawa}
  \state{Ontario}
  \country{Canada}
}

\author{Keheliya Gallaba}
\correspondingauthor
\orcid{0000-0002-5880-5114}
\email{gallabak@sigsoft.org}
\affiliation{%
  \institution{Centre for Software Excellence, Huawei Canada}
  \state{ON}
  \country{Canada}
  }

%commands
\newcommand{\newtext}[1]{\textcolor{black}{#1}}
\newcommand{\newt}[1]{\textcolor{black}{#1}}
\newcommand{\reftype}[1]{{\it #1}}

\newcommand{\Biruk}[1]{\textcolor{purple}{{\it [Biruk: #1]}}}
\newcommand{\todo}[1]{\textcolor{blue}{{\it [Biruk: #1]}}}
\newcommand{\etal}{et~al.}
\newcommand{\todoyiwei}[1]{\textcolor{orange}{{\it [yiwei: #1]}}}
\newcommand{\gmm}{\texttt{GMM}}
\newcommand{\hdbscan}{\texttt{HDBSCAN}}
\newcommand{\proposedsi}{\texttt{CASPER-SI}}
\newcommand{\casper}{\texttt{CASPER}}
%\setlength{\textfloatsep}{1pt}


\begin{abstract}
  
Regression testing for LLM-based systems poses unique challenges because individual regression instances provide limited information about system-level regressions. A failure in a single instance does not necessarily indicate a meaningful regression or provide sufficient information to diagnose affected behaviors. Conversely, detecting regressions based only on overall system performance changes is too coarse-grained, as it does not identify which behaviors are affected. This motivates analyzing regression instances at an intermediate level through test suite slices.

To address this challenge, we propose \casper{}, a change-aware slice prioritization framework for efficient regression testing of prompt-level and model-level changes in LLM-based systems. \casper{} first identifies regression slices containing semantically related instances with consistent performance characteristics using an evolutionary slice identification approach. Given a change to an LLM-based application, \casper{} prioritizes slices according to their likelihood of regression using behavioral information extracted from execution logs. We instantiate \casper{} in the software issue resolution domain and evaluate slice identification against clustering-based baselines and regressed slice prioritization against a random ranking baseline. Results show that \casper{} generates more consistent slices while maintaining comparable or better semantic coherence and improves regressed slice prioritization across different LLM-based system changes and testing budgets.
\end{abstract}




\maketitle

\section{Introduction}
Incorporating Large Language Models (LLMs) into software systems requires significant adjustments to traditional regression testing methodologies. An LLM-based system consists of an LLM integrated with a surrounding software scaffold that manages its interaction with the external environment. The scaffold executes the actions proposed by the LLM, collects the resulting observations, and feeds them back to the model, enabling iterative reasoning and action toward accomplishing a given task.

A regression instance consists of a pair of inputs and expected outputs. A regression test suite contains many regression instances, and a failure in any of them indicates a regression fault in the system under test (SUT).  In LLM-based systems, a single regression instance consists of a prompt and the expected textual output. The prompt is a natural-language description that includes instructions for the LLM on what to do, along with the input data it needs to process (e.g., user inputs or inputs from other system components). A test suite contains multiple regression instances that evaluate the LLM across dimensions such as factuality, toxicity, and regulatory compliance, with each regression instance using different input data and possibly different prompt instructions.

Regression tests are triggered by a change. In LLM-based systems, changes can occur at multiple levels of the application stack, including the underlying foundation model, system and user prompts, inference hyperparameters, supporting deterministic logic, and external tools or input/output processing components such as validation and filtering. Any of these changes may alter the performance of the LLM system, potentially introducing regressions or improvements in its tasks. Consequently, regression testing is required to identify regression instances affected by such changes while minimizing the cost of re-executing the regression test suite, an increasingly important requirement in CI/CD pipelines where changes are integrated and deployed continuously.

Unlike traditional software regression tests, where a failure on a single regression instance is interpreted as a fault in the software under test, the same definition of failure does not apply to software systems that contain ML models like LLMs. 
%since the model versions could have a better overall performance while mispredicting or generating new incorrect outputs in a few regression instances \todoyiwei{this sentence is confusing to me, why would ``better overall performance'' justify that per-case failure is wrong? Maybe something like this: 
Unlike traditional software, where a single failing regression test reliably indicates a fault, ML-based systems require a broader definition of failure. Because model outputs are probabilistic, a new version can improve overall accuracy while still mispredicting cases it previously got right. 
On the other hand, if we consider the LLM’s performance across the entire test suite, the results will be overly general and not informative for developers. One recommended approach in the literature is to divide the test suite into multiple slices, each containing multiple regression instances \cite{ma2024my}. The regression test fails when the LLM's performance falls below a threshold (e.g., the previous version's) for at least one slice. However, slicing the test suite is challenging, as the slices must be carefully selected to ensure that the regression instances are semantically related and that the SUT performs consistently on the majority of the regression instances within each slice. Proposing automated slicing techniques for this purpose is an open research problem.

%A variety of slice identification methods have been proposed, from decision-tree and lattice–based approaches in structured data\cite{chung2019slice}, to embedding-driven slice identification in images and time-series\cite{eyuboglu2022domino, ghosh2025ladder}, frequent-pattern-based techniques leveraging metadata\cite{zhang2022sliceteller}, and clustering-based methods explicitly designed for classification models\cite{hua2022discover, olesen2024slicing}. However, these techniques are typically modality- or task-specific, often relying on image encoders, classification outputs, or predefined corpora of slice descriptions. In contrast, we propose an approach that generalizes slice identification to LLM-based systems, including generative tasks, without requiring external description corpora or domain-specific model architectures, thereby enabling flexible, model-agnostic slicing suited for regression testing.

A variety of slice identification methods have been proposed, ranging from decision-tree- and lattice-based approaches for structured data~\cite{chung2019slice}, to embedding-driven methods for images and time-series~\cite{eyuboglu2022domino, ghosh2025ladder}, frequent-pattern-based techniques leveraging metadata~\cite{zhang2022sliceteller}, and clustering-based methods designed for classification models~\cite{hua2022discover, olesen2024slicing}. However, these techniques are primarily intended for analyzing or evaluating predictive models and are typically modality- or task-specific, often relying on image encoders, classification outputs, or predefined corpora of slice descriptions. To the best of our knowledge, they have not been applied to identify semantically coherent slices of regression instances for the regression testing of LLM-based systems. To address this gap, we propose a model-agnostic slice identification approach for LLM-based systems, including generative tasks, that does not require external description corpora or domain-specific model architectures.

\subsection{Application domain}

While the proposed approach can be generalized to any LLM system domain with customization or redesign of features used for ranking slices, regression dataset and performance metrics, we focus on the software engineering application domain for demonstration. In particular, we are focusing on automated code issue resolution using an LLM. The LLM application under test (LLMUT) is given an issue description and a codebase, and is tasked with analyzing the issue and generating a patch to fix it. The test suite contains multiple regression instances, each with a prompt, a gold patch, and the test cases that the generated patch must pass to resolve the issue. The prompt contains all the issue descriptions and associated information, such as comments and some code context from the codebase. The performance of the generated patch can be evaluated by metrics such as the test case pass rate. 

\subsection{Problem Definition}
\label{subsec:probst}
Given a regression testing dataset containing $N$ regression instances for a large language model application under test (LLMUT), the objective is to support cost-effective regression testing through two sequential tasks:

\begin{enumerate}
    \item \textbf{Slice Identification:} Partition the regression testing dataset into regression slices such that each slice
    \begin{enumerate}
    \item contains semantically coherent regression instances~\cite{eyuboglu2022domino}, so that regression instances within a slice share similar behavioral characteristics and can be represented by a small number of samples; and
    \item exhibits consistent LLMUT performance~\cite{eyuboglu2022domino}, where the regression instances within a slice are either predominantly successful or predominantly unsuccessful, enabling the execution results of representative regression instances to generalize to the remaining instances in the slice.
\end{enumerate}

    \item \textbf{Regressed Slice Ranking:} Given the identified regression slices and a modified version of the LLMUT, rank the slices according to their likelihood of being regressed by the change, thereby prioritizing slice execution under a limited testing budget.
\end{enumerate}

To address this problem, we propose \textbf{CASPER} (\textbf{C}hange-\textbf{A}ware \textbf{S}lice \textbf{P}rioritization for \textbf{E}fficient \textbf{R}egression Testing of LLM-based applications), a two-stage framework for cost-effective regression testing of LLM-based applications. In the first stage, the slice identification component of \casper{} (\proposedsi{}) identifies regression slices by partitioning the regression testing dataset into slices that satisfy the requirements defined above, using the execution results of the original system version 
(\texttt{V0}), before the change is applied. In the second stage, \casper{} learns a failure prediction model from \emph{behavioral signals}, i.e., quantitative features extracted from the agent's conversation logs that characterize its reasoning and execution behavior on the original system version (\texttt{V0}). 
After executing representative regression instances from each slice on the modified version (\texttt{V1}), \casper{} extracts the behavioral signals for each executed regression instance, estimates the change in its failure impact score between versions, and aggregates these score changes within each slice to rank the regression slices according to their likelihood of being impacted by the change.

We instantiated \casper{} in the software issue resolution domain and evaluated both the overall framework and its slice identification component. We evaluated \casper{} against a random slice ranking baseline to assess its ability to prioritize regressed slices under limited testing budgets. In addition, we evaluated the slice identification component (\proposedsi{}) against two clustering-based baselines, \gmm{} and \hdbscan{}, to assess the quality of the generated regression slices.

The evaluation results demonstrate that \proposedsi{} effectively identifies high-quality regression slices by producing slices that contain semantically coherent regression instances and exhibit consistent LLMUT performance. Compared with clustering-based baselines, \proposedsi{} achieves substantially more consistent LLMUT performance within slices while maintaining comparable or better slice coherence. Furthermore, \casper{} consistently and significantly improves the prioritization of regressed slices compared to random ranking, achieving higher ranking effectiveness across all evaluated datasets and testing budgets. These improvements are particularly important under limited testing budgets, where only a small fraction of slices can be executed, and early identification of regressed slices is critical, especially in CI/CD pipelines where frequent changes require efficient regression testing to maintain rapid development and deployment cycles.

To summarize, the key contributions of the paper are as follows:
\begin{itemize}
    \item \textbf{CASPER:} An end-to-end framework that identifies meaningful regression slices and prioritizes regressed slices after LLM-based system changes, enabling cost-effective regression testing of LLM-based systems under limited execution budgets.

    \item \textbf{An evolutionary slice identification approach, \proposedsi{}:} Formulates slice identification as a multi-objective optimization problem and employs NSGA-II to generate regression slices that satisfy requirements of performance consistency and coherence.

    \item \textbf{A behavioral signal-based impacted slice ranking approach:} Learns a failure prediction model from behavioral signals extracted from LLM agent execution logs and uses changes in predicted failure impact between system versions to rank regression slices according to their likelihood of being affected by a change.

    \item \textbf{Comprehensive empirical evaluation in the software issue resolution domain:} Evaluates CASPER against clustering-based slice identification baselines and a random slice ranking baseline, demonstrating significant improvements in slice quality and more effective regressed slice prioritization across different LLM-based system changes.
\end{itemize}

The remainder of this paper is organized as follows: Section \ref{sec:approach} discusses the details of the proposed slice identification and description approach. Section \ref{sec:eval} outlines the case study and evaluation details. Section \ref{sec:result} presents the results of the evaluation case study. Section \ref{sec:threats} outlines threats to validity. Section \ref{sec:related} discusses related work, and finally Section \ref{sec:conclusion} concludes the paper.

\section{Proposed approach}
\label{sec:approach}
\begin{figure*}[!htbp]
\centering
\includegraphics[width=\linewidth]{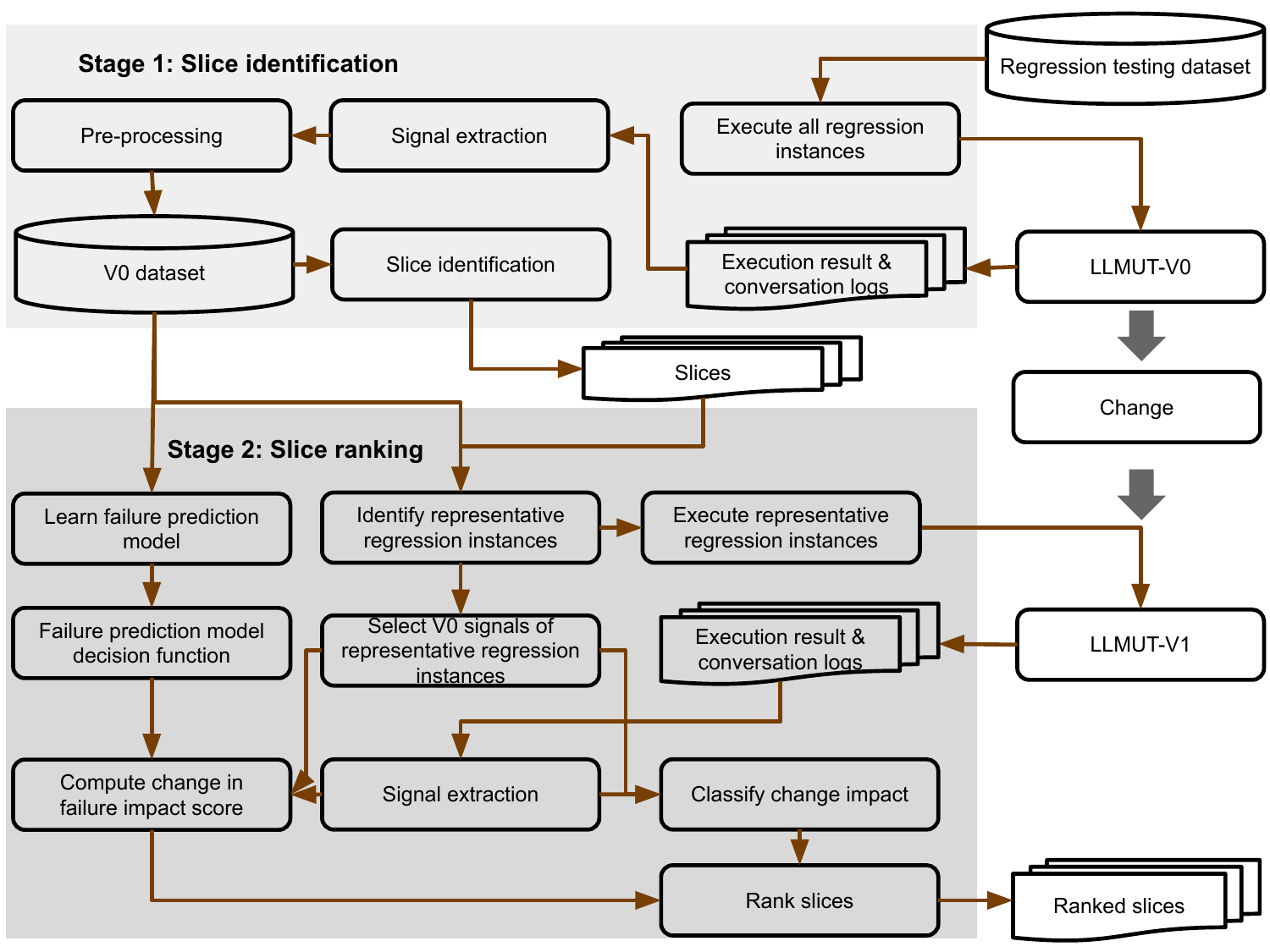}
\caption{Overview of \casper{}. A regression testing dataset is first executed on the LLMUT (\texttt{V0}) to collect execution result and conversation logs, from which behavioral signals are extracted. Stage 1 identifies slices, while Stage 2 ranks the slices according to their likelihood of being impacted after a system change, producing a prioritized list of slices.}
\Description{Overview of \casper{}. A regression testing dataset is first executed on the LLMUT (\texttt{V0}) to collect execution result and conversation logs, from which behavioral signals are extracted. Stage 1 identifies slices, while Stage 2 ranks the slices according to their likelihood of being impacted after a system change, producing a prioritized list of slices.}

\label{fig:Approach}
\end{figure*}

In this section, we describe \casper{}, a proposed approach to rank slices based on regression impact scores given a change in the LLMUT. Figure \ref{fig:Approach} shows the overview of our approach. \casper{} contains two stages that are Slice-identification and slice ranking.

\subsection{Inputs for \casper{}}

The input to the proposed approach consists of the execution results and LLMUT conversation logs obtained by running the regression testing dataset on the LLMUT original version (LLMUT-V0). The agent conversation logs contain the exchanged messages between the reasoning model and its environment. The reasoning model generates it's thoughts (reasoning steps) and actions to execute in the environment and obtains feedback from the environment on the results of those actions. The execution result depends on the task assigned to the LLMUT. For example, if the task of the LLMUT is to resolve software issues, given issue problem statements and a codebase as context, the LLMUT typically generates a patch to address the given issue. The generated patch is evaluated for correctness against an oracle, and the execution result is the oracle's evaluation of the patch. In this case, the generated patch will be evaluated as successful if it passes all test cases associated with the issue, and as unsuccessful otherwise. The execution result will be the evaluation result for a regression instance obtained from an oracle. For the software issue resolution domain, we define the \texttt{\textit{Patch success}} variable to capture the evaluation result, which is a binary variable whose value is either $1$ if the patch passes all test cases or $0$ otherwise.

\subsection{Stage 1: Slice identification (\proposedsi{})}

Stage 1, the slice identification stage, aims to identify subsets of regression instances where the LLMUT has consistent performance while occupying compact regions in the behavioral signal space. This stage consists of three main steps: behavioral signal extraction, signal preprocessing, and multiobjective search-based slice identification, which are described in the following subsections.
\subsubsection{Extraction of behavioral signals}
\label{subsec:signal-ref}
The first step of the slice identification is the extraction of behavioral signals. Since the prediction of regressed slices relies on this step, we describe it in detail. 

Behavioral signals capture or measure how the LLMUT's effort is distributed across the steps required to successfully perform the assigned task. Behavioral signals are domain-specific, but they generally reflect how well the LLMUT's actions align with the task objectives. For our application domain, software issue resolution, the LLMUT typically needs to first understand the problem statement, explore the code base context and environment, plan a fix strategy, implement the fix strategy by editing the code base, test the changes if they resolve the issue, refine based on test execution results, and finally generate a patch representing all applied changes to the code base. Behavioral signals are computed by parsing agent conversation logs.

Our approach assumes access to agent interaction logs containing the sequence of agent actions, tool invocations, tool outputs, and execution feedback. While such logs are not guaranteed to be available in every deployed agent system, maintaining interaction traces is a common practice for agent monitoring, debugging, auditing, and performance evaluation. Existing agent frameworks typically provide mechanisms to record these interactions, as they are essential for understanding agent behavior and diagnosing failures. Therefore, our approach requires that the agent execution environment exposes these interaction traces; it does not assume access to any domain-specific information beyond the logged agent trajectory.

In a typical ReAct agent configuration \cite{yao2022react}, the reasoning model generates a reasoning step and a corresponding action, which may involve invoking an external tool. The output of the tool call, including any returned results or errors, is then provided back to the model as feedback. The model uses this feedback to reason about the current state and determine its next action. Since the structure of conversation logs varies across agent implementations, extracting these interactions requires a custom parser for each agent. Here, we focus on describing the information required to extract the behavioral signals.

For the software issue resolution application domain, we require conversation log parsers that process every turn in an agent's execution trace. For each turn, the parser must extract the model's thought section and the corresponding turn number. Additionally, each turn must be mapped to one of the agent states described in Table \ref{tab:agent-state}. This table contains the agent state definition and its numerical encoding, which maintains logical order. We extended the agent states defined in \cite{liu2026evaluating} to include the \texttt{SETUP, CLEANUP, and SUBMISSION} states, which correspond to the initial setup, final cleanup, and submission of the generated patch, respectively. The \texttt{OTHER} state is used for turns that do not correspond to any of the defined states. We assigned a numerical encoding of -1 to the \texttt{OTHER} state to indicate that it does not belong to the main workflow states.

For turns that involve a tool call, the parser must also extract the tool call command, the tool execution result, and any associated error information (if present). The output of the conversation log parser is a sequence of structured records, each one corresponding to a single turn and containing the extracted thought section, turn number, agent state, and, when applicable, the tool call command, execution result, return code, and error details.

\begin{table}[]
\caption{Software issue resolution agent states.}
\label{tab:agent-state}

\begin{tabular}{@{}lp{8cm}l@{}}
\toprule
{State} & { State definition}                                                                     & \textbf{State order encoding} \\ \midrule
SETUP  & The agent performs the environment setup and installation of dependencies required to solve and test the given task & 0 \\
NAVIGATION                        & The agent performs exploration of the codebase and environment to understand the given task. It also examines a specific file or module in a codebase associated with the task at hand  & 1                             \\
REPRODUCTION                      &The agent attempts to reproduce the reported issue by executing the relevant program, running failing test cases, or constructing a minimal failing example to confirm and understand the given issue.   & 2                             \\
PATCH                       & The agent performs code editing, adding new code or adding a new file to execute the task       & 3                             \\
VALIDATION                       & The agent tests if the applied change resolved the given task or not                          & 4                             \\

CLEANUP & The agent cleans up temporary files and release resources that were used for patch generation.                      & 5
\\
SUBMISSION                       & The agent submits the final patch that solves the given issue                                & 6                             
\\ OTHER & The agent performs other actions that are not part of the main workflow states. &  \\ \bottomrule
\end{tabular}%

\end{table} 

The following paragraphs describe the signal extraction component of \casper{}, including how each behavioral signal is computed from the parsed conversation logs and execution results.

$\blacksquare$ \texttt{File-level recall:}

We compute \texttt{File-level recall} to capture the alignment between the agent's generated patch and the gold patch. 

The \texttt{file-level recall} is computed by counting the number of files that are edited in both the generated patch and the gold patch and dividing this count by the total number of files edited in the gold patch. A value closer to 1 indicates that the generated patch and the gold patch modify similar files, suggesting convergence of the solution.

The rationale for this signal is that stronger file-level alignment between the generated and gold patches increases the likelihood that the generated patch is correct. Since issue-resolution regression datasets are collected from historically resolved issues, they include the accepted repair patch for each regression instance. Although a gold patch is not required to determine patch correctness when an executable test suite is available, it provides additional information that can be used to compute \texttt{file-level recall}.

$\blacksquare$ \texttt{Mean absolute state movement (MASM):}

The signal measures the overall magnitude of transitions between execution states by representing each state except \texttt{OTHER} using the encoding defined in Table \ref{tab:agent-state}. The state \texttt{OTHER} is excluded here because no logical order can be assigned to it. MASM is computed using Equation \ref{eq:total_movement}, where S(t) denotes the state encoding at turn t, and T is the total number of turns.

\begin{equation}
\label{eq:total_movement}
\text{MASM} =
\frac{\sum_{t=1}^{T} \left| S(t) - S(t-1) \right|}{T - 1}
\end{equation}

MASM captures the stability of the agent’s state trajectory. Low values indicate stagnation in a small subset of states, while high values indicate oscillatory transitions across states without convergence. Effective trajectories are expected to exhibit moderate MASM values, reflecting steady progression toward a solution.

$\blacksquare$ \texttt{Trajectory Entropy Deviation (TED):}

This signal measures how far the agent's state-transition trajectory entropy deviates from a baseline distribution derived from state-transition trajectories that led to successful patch generation. TED is computed using Equation~\ref{eq:entropy_deviation}, where (H) denotes the Shannon entropy computed from the empirical distribution of state encodings, in which the probability of each state is estimated as its relative frequency within the trajectory. The terms ($\mu_H$) and ($\sigma_H$) denote the mean and standard deviation, respectively, of the entropy values computed over successful trajectories only. The z-score transformation is applied to quantify the deviation of a trajectory's entropy from the typical entropy observed among successful patch generations. This standardization provides a scale-independent measure of whether a trajectory exhibits unusually high or low state-transition variability relative to successful trajectories. A TED value close to zero indicates entropy behavior consistent with successful trajectories, whereas larger positive or negative values indicate greater deviation from this baseline.

\begin{equation}
\label{eq:entropy_deviation}
\text{TED} = \frac{H - \mu_H}{\sigma_H}
\end{equation}

A higher TED value indicates that the trajectory exhibits a higher state-distribution entropy than typical successful trajectories, suggesting a deviation from successful execution patterns. 

The rationale for this signal is that trajectories with higher state-distribution entropy than successful trajectories are more likely to result in incorrect or incomplete patch generation.

$\blacksquare$ \texttt{Testing effort:} 

This signal measures the effort the agent devotes to validating candidate patches. We compute the testing effort as the proportion of conversation turns assigned to the \texttt{VALIDATION} state, i.e., the number of \texttt{VALIDATION} turns divided by the total number of turns. This signal reflects how much of the agent's interaction is dedicated to validation. A relatively high value may indicate thorough validation before producing a correct patch, but it may also reflect repeated validation attempts caused by persistent failures. Conversely, a relatively low value may indicate that the agent reached a correct solution with little validation, or that it terminated with insufficient validation. Therefore, this signal captures deviations in validation behavior that may distinguish successful patches from incorrect ones.

$\blacksquare$ \texttt{Test failure rate:} 

Test failure rate measures the proportion of tests that resulted in an error. We compute the test failure rate by first counting turns whose state is \texttt{VALIDATION} and whose corresponding return code is nonzero (indicating a test error). Next, we divide this count by the number of turns spent for validation. The value of this signal ranges from 0 to 1, where a value closer to 1 indicates the agent is succeeding in most validations, and a value closer to 0 indicates the agent is struggling with failed validations, which could ultimately lead to generating incorrect patches or failing to generate any patches.

$\blacksquare$ \texttt{Dominant File Ratio:}

This signal measures the extent to which the agent repeatedly modifies the same file during issue resolution. We first extract the edited files from all code editing actions performed during the \texttt{PATCH} state in the agent's conversation logs. We count each occurrence of a file in the extracted set of edited files. Consequently, if a file is modified multiple times during execution, each occurrence contributes to the overall count.

Let $c_f$ denote the number of occurrences of file $f$ across all extracted edited files lists, and let $F$ denote the set of distinct edited files. The dominant file ratio is defined as

\[
\mathrm{DFR}
=
\frac{\max_{f \in F} c_f}
{\sum_{f \in F} c_f}.
\]

The numerator represents the number of occurrences of the most frequently modified file, while the denominator represents the total number of file occurrences across all code editing actions. A value close to 1 indicates that the agent repeatedly modifies the same file throughout the execution, whereas a lower value indicates that editing activity is distributed across multiple files.

Software issue resolution is often characterized by localized faults, where successful patches typically modify a small number of files closely related to the root cause. Consequently, a higher dominant file ratio suggests more focused editing behavior and potentially better fault localization, while a lower ratio may indicate broader exploration, unnecessary modifications across multiple files, or weaker localization, all of which may correlate with lower patch success.

$\blacksquare$ \texttt{Cross-state file overlap (CSFO):} measures the consistency of the agent's affected files across different workflow states. It quantifies the extent to which the same files are involved during navigation, editing, and testing.

Let $S_i$ denote the set of affected files in state $i$, where $i \in \{\text{Navigation}, \text{Patch}, \text{Validation}\}$. The similarity between two states $i$ and $j$ is measured using the Jaccard index \cite{leskovec2020mining}:

\[
J_{ij} = \frac{|S_i \cap S_j|}{|S_i \cup S_j|}
\]

The cross-state file overlap is defined as the average Jaccard similarity across all pairs of workflow states:

\[
\text{CSFO}
=
\frac{1}{3}
\sum_{i\neq j} J_{ij},
\]

Higher values indicate that similar sets of files are involved across workflow states, reflecting a consistent debugging focus throughout the agent's trajectory. Lower values indicate that different files are involved at different workflow states, suggesting greater shifts in the agent's focus during debugging.

 $\blacksquare$ \texttt{Plan compliance signals:}

   Liu \etal~\cite{liu2026evaluating} proposed three plan compliance metrics to measure the alignment between an instruction-level plan and an agent's actual execution. An instruction-level plan is represented as an ordered sequence of agent states, where each state corresponds to a distinct phase of the issue resolution process inferred from the instruction prompt. For example, a plan may consist of the sequence \texttt{{SETUP, NAVIGATION, REPRODUCTION, PATCH, VALIDATION, CLEANUP, SUBMISSION}}. The proposed metrics are: \texttt{Plan Phase Compliance (PPC)}, \texttt{Plan Order Compliance (POC)}, and \texttt{Plan Phase Fidelity (PPF)}. Based on an empirical study of software issue resolution agent execution traces, Liu \etal~\cite{liu2026evaluating} found that higher compliance with the instructed plan is positively correlated with successful patch generation. Motivated by this finding, we include these three metrics as behavioral signals in our approach.

    The detailed definitions and computation procedures of these metrics are provided in the original paper~\cite{liu2026evaluating}. For completeness and to make the paper self-contained, we briefly summarize them below.

\texttt{Plan Phase Compliance (PPC)~\cite{liu2026evaluating}}: Let $P$ denote the set of states in the given plan, and let $E$ denote the set of states observed in the agent's execution. The plan phase compliance is defined as the ratio of the intersection of $P$ and $E$ to the total number of phases in the plan:

    \begin{equation}
    \mathrm{PPC}
    =
    \frac{\left| P \cap E \right|}{\left| P \right|}
    \end{equation}

\texttt{ Plan Order Compliance (POC)~\cite{liu2026evaluating}:}
Let $O_P$ denote the ordered sequence of states in the given plan, and let $O_E$ denote the ordered sequence of the first occurrences of each state in the agent's execution. The plan order compliance measures the extent to which the ordering of the agent's states aligns with the given plan's ordering. It is defined as the ratio of the length of the longest common subsequence (LCS) between $O_P$ and $O_E$ to the total number of states in the plan:

\begin{equation}
\mathrm{POC}
=
\frac{\left|\mathrm{LCS}(O_P, O_E)\right|}{|O_P|}.
\end{equation}

\texttt{ Plan Phase Fidelity (PPF)~\cite{liu2026evaluating}:} 

Agents may adjust execution based on the current context rather than the initial plan, which may leave the agent in a state not part of the given plan. While introducing new states by itself is not inherently negative, it may distract the agent in subsequent steps and lead it to produce incorrect or incomplete patches. Let $P$ denote the set of states in the given plan, and let $E$ denote the set of states observed in the agent's execution. The Plan Phase Fidelity (PPF) penalizes the presence of states in the agent's execution that fall outside the specified plan states. It is defined as the ratio of the number of elements of $P$ and $P \cup E$:
\begin{equation}
\mathrm{PPF}
=
\frac{|P|}
{|P \cup E|}
\end{equation}

\subsubsection{Preprocessing} We first compute the 10 behavioral signals described previously from the agent conversation logs. The resulting features are then standardized using z-score normalization (zero mean and unit variance) to ensure that all signals are on a comparable scale across issues. This preprocessing step produces the \texttt{V0 dataset} (see Figure \ref{fig:Approach}), which serves as the input to \proposedsi{} and slice ranking. The \texttt{V0 dataset} contains the standardized behavioral signal values for each instance in the regression testing dataset, together with the corresponding \texttt{patch success} label.

\subsubsection{Slice identification using Multi-objective search}
In the context of automated software issue resolution using agentic systems, a \emph{slice} is intended to group issues that exhibit similar behavioral characteristics with respect to the LLMUT. Specifically, a desirable slice should contain issues for which the LLMUT demonstrates consistent performance \cite{eyuboglu2022domino}, either by consistently generating \emph{good patches} that pass all test cases or by consistently producing \emph{unacceptable patches} that fail at least one test case. Such behavioral consistency enables reliable characterization of the LLMUT's strengths and weaknesses on a subset of issues. Furthermore, a slice should exhibit high behavioral coherence \cite{eyuboglu2022domino}. Specifically, issues within a slice should have similar behavioral signal vectors, resulting in small intra-slice distances and large inter-slice distances in the behavioral signal space. High behavioral coherence improves the interpretability of a slice and increases the likelihood that the observed performance patterns arise from shared underlying issue characteristics rather than incidental similarities.

These desired slice properties naturally define competing optimization objectives. For example, maximizing output consistency may not always maximize behavioral coherence, and vice versa. Consequently, the slice identification problem can be formulated as a multi-objective optimization problem in which candidate slices are evaluated according to these quality properties.

We tackle the slice identification problem  using multi-
objective search algorithms, where the desired slice quality properties are treated as objectives to optimize. Several multi-objective search techniques can be considered, including: \textit{Scalarization} \cite{Luke2013Metaheuristics}, which converts multiple objectives into a single objective using a weighted sum utility; \textit{Evolutionary algorithms} \cite{deb2011multi}, which employ well-known population-based search and evolutionary techniques to generate a diverse set of solutions (e.g., NSGA-II, NSGA-III \cite{deb2013evolutionary}, MOSA \cite{amine2019multiobjective}); and \textit{Indicator-based approaches} \cite{zitzler2004indicator}, which use performance indicators to guide environmental selection, evaluating the contribution or quality of candidate solutions with respect to the current population rather than relying solely on Pareto dominance or decomposition.  

Among the multi-objective search techniques, we resorted to NSGA-II for slice identification for the following reasons:
\begin{enumerate}
    \item NSGA-II is efficient for problems with fewer than four objectives, maintaining solution diversity using crowding distance.
   \item NSGA-II is well-supported, with a solid theoretical foundation and robust implementation libraries.
\end{enumerate}

Evolutionary-based multi-objective search techniques involve decisions about population encoding, population initialization, and objective evaluation, followed by genetic operations on individuals. The details of each step are outlined below.

$\blacksquare$ \texttt{Population encoding}

Evolutionary multi-objective search techniques operate on a population of candidate solutions by first evaluating fitness and then applying genetic variation to drive optimization. Individuals in the population compete for selection based on their fitness evaluations.

Each individual is represented by a list of $N$ slice IDs, where $N$ is the number of regression instances. The indices correspond to regression instances in the input dataset, and the values indicate the slice to which each instance belongs. For example, with $N=4$, the individual $I=[0,0,1,2]$ defines three slices: slice 0 contains $[t_0, t_1]$, slice 1 contains $[t_2]$, and slice 2 contains $[t_3]$. Slice sizes can vary across individuals. Although this representation requires custom initialization, mutation, and crossover operators, it enables efficient, scalable handling of large test suites by allowing direct indexing of regression instances into slices, facilitating vectorized computations, and supporting flexible slice sizes with minimal memory overhead.

$\blacksquare$ \texttt{Population initialization.}

The population initialization component is parameterized by the hyperparameters \texttt{number of individuals}, \texttt{minimum slice size}, and \texttt{maximum slice size}. During this stage, individuals are randomly initialized while ensuring that the slice size constraints are respected, that each regression instance belongs to exactly one slice, and that all regression instances are covered. The number of slices per individual may vary across the population. 

The minimum and maximum slice sizes are important parameters in regression testing. Larger slice sizes make the regression analysis coarse-grained, reducing its usefulness to developers by obscuring fine-grained distinctions between regression instances. Conversely, very small slice sizes yield trivial partitions and make regression analysis intractable, producing a large number of tiny slices that overwhelm downstream analysis.

$\blacksquare$ \texttt{Objective functions and constraints.}

The optimization is guided by multiple objectives that encapsulate the desired properties of high-quality slices, namely: \texttt{output-inconsistency (OI), Intra-slice cohesion, Inter-slice distance}. The details of each objective are described below.

\begin{enumerate}
    \item \texttt{Output Inconsistency (OI)}
    The output inconsistency objective measures the extent to which the LLM application under test (LLMUT) exhibits inconsistent performance within a slice. Since the notion of performance depends on the application domain, the metric is domain-specific. In the automated software issue resolution domain, performance is determined by the success of the generated patch. For an issue $i$, the patch success is defined as

        \begin{equation}
        PS(i)=
        \begin{cases}
        1, & \text{if the generated patch passes all test cases},\\
        0, & \text{otherwise.}
        \end{cases}
        \end{equation}

        For a slice $S$, let the dominant patch success value be the majority value among the issues in the slice:

        \begin{equation}
        d(S)=
        \arg\max_{v\in\{0,1\}}
        \left|\left\{ i \in S : PS(i)=v \right\}\right|.
        \end{equation}

        The output consistency of the slice is defined as the proportion of issues whose patch success agrees with the dominant value:

        \begin{equation}
            \label{eq:oc_metric}
        \mathrm{OutputConsistency}(S)
        =
        \frac{\left|\left\{ i \in S : PS(i)=d(S) \right\}\right|}{|S|}.
        \end{equation}

        The output inconsistency of the slice is then computed as

        \begin{equation}
        \mathrm{OutputInconsistency}(S)
        =
        1-\mathrm{OutputConsistency}(S),
        \end{equation}

        where lower values indicate more consistent LLMUT behavior within the slice.

       Finally, given an individual solution $i$ consisting of $m$ slices, the output inconsistency objective for individual $i$ is defined as the mean output inconsistency across all slices:

        \begin{equation}
        \mathrm{OI}(i)
        =
        \frac{1}{m}
        \sum_{j=1}^{m}
        \mathrm{OutputInconsistency}(S_j).
        \end{equation}

        The optimization objective is to \textit{minimize} $\mathrm{OI}(i)$.

    \item \texttt{Intra-Slice Distance (Intra\_SD).}
        The intra-slice distance objective measures the compactness of a slice in the behavioral signal space. Let each issue $i$ be represented by a behavioral signal vector $\mathbf{v}_i \in \mathbb{R}^d$. For a slice $S$, the intra-slice distance is defined as the average pairwise Euclidean distance between all issue representations within the slice:

        \begin{equation}
        \mathrm{Intra\_SD}(S)
        =
        \frac{2}{|S|(|S|-1)}
        \sum_{\substack{i,j \in S \\ i < j}}
        \|\mathbf{v}_i - \mathbf{v}_j\|_2.
        \end{equation}

        The objective is to minimize the mean intra-slice distance across all slices. For a solution consisting of $m$ slices, the individual-level objective is:

        \begin{equation}
        \mathrm{Intra\_SD}(i)
        =
        \frac{1}{m}
        \sum_{k=1}^{m}
        \mathrm{Intra\_SD}(S_k).
        \end{equation}

        The optimization objective is to \textit{minimize} $\mathrm{Intra\_SD}(i)$.
   \item \texttt{Inter-Slice Distance (Inter\_SD).}
        The inter-slice distance objective measures the separation between slices in the behavioral signal space. Each slice $S$ is represented by its centroid, defined as the mean of its behavioral signal vectors:

        \begin{equation}
        \boldsymbol{\mu}(S)
        =
        \frac{1}{|S|}
        \sum_{i \in S} \mathbf{v}_i.
        \end{equation}

        The inter-slice distance between two slices is defined as the Euclidean distance between their centroids:

        \begin{equation}
        d(S_a, S_b)
        =
        \|\boldsymbol{\mu}(S_a) - \boldsymbol{\mu}(S_b)\|_2.
        \end{equation}

        For a slice $S_a$, its separation score is defined as the minimum distance to any other slice:

        \begin{equation}
        \mathrm{Inter\_SD}(S_a)
        =
        \min_{b \neq a}
        d(S_a, S_b).
        \end{equation}

        The objective is to \textit{maximize} inter-slice separation. At the individual level, the objective is defined as the mean inter-slice distance across all slices:

        \begin{equation}
        \mathrm{Inter\_SD}(i)
        =
        \frac{1}{m}
        \sum_{k=1}^{m}
        \mathrm{InterCD}(S_k).
        \end{equation}

        The optimization objective is to maximize $\mathrm{Inter\_SD}(i)$.   
        \end{enumerate}

During each fitness evaluation, the minimum and maximum slice-size constraints are checked to determine whether the current individual is a valid solution or violates either constraint. 

$\blacksquare$ \texttt{Selection and genetic operations.}

We implemented custom crossover and mutation operators tailored to our individual representation, enabling the multi-objective evolutionary algorithm to effectively explore the search space while maintaining valid slice assignments.

Each individual is represented as a fixed-length vector of slice IDs, where the index corresponds to a regression instance and the value indicates its assigned slice. Therefore, standard permutation- or subset-based crossover operators are not applicable. We employ a customized vectorized crossover operator inspired by the Half Uniform Crossover (HUX) mechanism~\cite{eshelman1991chc}. The operator consists of identifying differences between parents, performing HUX-style swapping, and repairing slice-size constraint violations.

Given two parent vectors $p_1,p_2 \in \mathbb{Z}^N$, where $p_i[t]$ denotes the slice ID assigned to regression instance $t$, we first identify the positions where the parents disagree:
\[
D = \{\,t \mid p_1[t] \neq p_2[t]\,\}.
\]
Following the HUX principle, $\lfloor |D|/2 \rfloor$ positions are randomly selected from $D$ and swapped between the parents to generate two offspring:
\[
o_1[t],o_2[t] \leftarrow o_2[t],o_1[t].
\]
This operation introduces diversity while preserving the overall genome structure.

After crossover, offspring may violate the predefined slice-size constraints. For each slice ID $s$, we define the corresponding set of assigned regression instances as:
\[
S_s=\{\,t\mid o[t]=s\,\}.
\]
The repair operator ensures that every slice satisfies:
\[
\text{min\_size}\leq |S_s|\leq \text{max\_size}.
\]
Oversized slices are divided, while instances from undersized slices are reassigned to existing slices with available capacity.

For example, consider two parents with $N=6$ regression instances:
\[
p_1=[0,0,1,1,1,1], \qquad
p_2=[0,2,2,2,2,2],
\]
and constraints $\text{min\_size}=2$ and $\text{max\_size}=3$. The parents differ at:
\[
D=\{1,2,3,4,5\}.
\]
Since $|D|=5$, two positions are selected for swapping. Assuming positions $\{2,5\}$ are selected, the resulting offspring are:
\[
o_1=[0,0,2,1,1,2], \qquad
o_2=[0,2,1,2,2,1].
\]

The first offspring already satisfies the constraints because each slice contains two instances. For the second offspring, slice $S_0$ contains only one instance and violates the minimum size requirement:
\[
S_0=\{0\},\quad |S_0|=1.
\]
The invalid slice is removed and its instance is reassigned to the smallest valid slice:
\[
o_2=[1,2,1,2,2,1].
\]
The final slice sizes become:
\[
|S_1|=3,\qquad |S_2|=3,
\]
which satisfy the required constraints. Therefore, the proposed crossover operator preserves the vector representation, introduces controlled diversity through HUX-based swapping, and maintains valid slice partitions through constraint-aware repair.

The mutation operator follows the fixed-length slice-ID representation used by the evolutionary search. We implement a constraint-aware, vectorized mutation strategy that combines move and swap operations while preserving valid slice partitions. Each mutation step randomly determines the number of move and swap attempts, bounded by user-defined limits, and executes each attempt with an independent probability (\texttt{move\_prob} or \texttt{swap\_prob}). All operations are subject to the minimum and maximum slice-size constraints.

A move mutation selects a source slice $S_i$ satisfying
$\lvert S_i\rvert > \text{min\_slice\_size}$ and transfers a randomly selected regression instance to a destination slice $S_j$ satisfying
$\lvert S_j\rvert < \text{max\_slice\_size}$. This operation changes slice assignments while preventing empty or oversized slices. A swap mutation selects two distinct non-empty slices and exchanges one randomly selected regression instance from each. Since swap operations preserve slice sizes, they introduce structural diversity without violating partition feasibility.

After mutation, a repair procedure is applied to handle possible undersized slices. Specifically, instances belonging to slices with $\lvert S_i\rvert < \text{min\_slice\_size}$ are temporarily removed and reassigned to existing slices with available capacity. If no valid slice can accommodate the instances, a new slice ID is created. This guarantees that all resulting individuals satisfy the required slice-size constraints.

For example, consider an individual with $N=8$ regression instances:
\[
\mathbf{x}=[0,0,0,\;1,1,\;2,2,2],
\]
where the slice sizes are $|S_0|=3$, $|S_1|=2$, and $|S_2|=3$. Assume
$\text{min\_slice\_size}=2$ and $\text{max\_slice\_size}=4$. 

A move mutation first selects a valid source slice from:
\[
\{S_i \mid |S_i|>2\}=\{0,2\}.
\]
Assuming $S_0$ is selected and the element at position 1 is moved, valid destination slices are:
\[
\{S_j \mid |S_j|<4\}=\{1,2\}.
\]
Selecting $S_1$ as the destination produces:
\[
\mathbf{x}=[0,\mathbf{1},0,\;1,1,\;2,2,2].
\]

A subsequent swap mutation selects two slices, for example $S_1$ and $S_2$, and exchanges one element from each:
\[
S_1=\{1,3,4\},\qquad S_2=\{5,6,7\}.
\]
Swapping positions 3 and 6 results in:
\[
\mathbf{x}=[0,1,0,\;2,1,\;2,1,2].
\]
The resulting slice sizes remain valid:
\[
|S_0|=2,\qquad |S_1|=3,\qquad |S_2|=3.
\]
Since no slice violates the minimum size constraint, the repair phase is not required. Therefore, the final mutated individual is:
\[
\mathbf{x}^{\prime}=[0,1,0,\;2,1,\;2,1,2].
\]

$\blacksquare$ \texttt{Post-processing and knee-point selection.}

The optimization process produces a set of Pareto-optimal individuals (Pareto front). From this set, the best individual is selected using a knee-point analysis of the Pareto front. A \emph{knee} region is characterized by solutions in which a small improvement in one objective leads to a significant deterioration in another; such points typically represent strong compromises across all objectives.

If multiple knee points are detected, the point closest to the origin in the objective space is selected. Since all objectives are formulated as minimization objectives, the ideal point corresponds to the origin. To determine the closest knee point, we compute the magnitude (Euclidean norm) of each knee point and choose the individual with the smallest magnitude as the final solution for the slice identification process.

$\blacksquare$ \texttt{Slice identification hyper-parameters.}

 We describe the hyperparameters of the slice identification algorithm as follows.
\begin{enumerate}
    \item \texttt{Population\_size:} Specifies the number of individuals in the initial population and the maximum population size maintained throughout the optimization process. This parameter influences the search diversity and convergence behavior.

    \item \texttt{Crossover\_prob:} Defines the probability of applying crossover during the evolution operation. This controls the likelihood of combining two parent individuals to generate offspring.

    \item \texttt{Mutation\_prob:} Defines the probability of applying a mutation to an individual. This value governs the likelihood of introducing random variations, thereby increasing exploration in the search space.
    \item \texttt{Max\_mutations:} Defines the maximum number of move and swap operations that may be applied during a single mutation step.
    \item \texttt{Max\_generations:} Specifies the total number of generations (iterations) the evolutionary algorithm performs. This controls the computational budget and the depth of the search.

    \item \texttt{Min\_slice\_size:} Imposes a constraint on the minimum number of regression instances that each slice must contain throughout the optimization process. The value must lie within the interval $[2, N]$, where $N$ is the total number of regression instances in the test suite. This ensures that slices remain sufficiently informative while avoiding excessively fine-grained slices.
   \item \texttt{Max\_slice\_size:} Imposes an upper bound on the number of regression instances that a slice may contain during the optimization. Its value must lie within $(Min\_slice\_size, N]$, where $N$ denotes the total number of regression instances in the suite. This constraint discourages the creation of excessively large slices.

\end{enumerate}

\subsection{Stage 2: Slice ranking (\proposedsi{})}
The objective of Stage 2 is to estimate the regression impact of each slice following a change to the LLMUT. To accomplish this, we first learn a failure prediction model from the original regression testing dataset and then use it to assess how the predicted failure impact of representative regression instances changes after executing them on the updated LLMUT. The following subsections describe the learning, execution, signal extraction, and impact ranking procedures in detail.
\subsubsection{Learn failure prediction models on the original dataset}
In this stage, we leverage the \texttt{V0 dataset} to learn a failure prediction model that maps behavioral signals to the corresponding task performance metric, namely \texttt{patch success} in the software issue resolution domain. While a variety of machine learning models could be employed for this task, including random forests, gradient boosting, and neural networks, our preliminary experiments showed that \texttt{Logistic Regression} and \texttt{Support Vector Machines (SVM)} provided the best performance. Therefore, we use these two models in our study.

The selected model is trained using the \texttt{V0 dataset} to learn the relationship between behavioral signals and patch success. After training, each regression instance in the \texttt{V0 dataset} is assigned a failure impact score using the model's decision function. For Logistic Regression, the failure impact score is the predicted probability of patch success. For SVM, it is the signed distance of the regression instance from the separating hyperplane. These scores quantify the model's confidence in its prediction and are subsequently used to rank slices by their estimated regression impact.

\subsubsection{Identify representative regression instance(s)}
Given the number of samples per slice $K$, we select $K$ regression instances for a slice for execution on LLMUT-V1 (LLMUT after a change is applied) using two sampling approaches: \texttt{Closest distance sampling (CDS)} and \texttt{Furthest point sampling (FPS)}. The former approach selects the $K$ regression instances closest to the centroid of a slice, while the latter focuses more on diversity and selects $K$ regression instances that are maximally separated in the behavioral signal space. The details of each sampling approach are described below. 

\paragraph{\texttt{Closest distance sampling}}

Given a slice $s$, let $X_0[s] \in \mathbb{R}^{|s|\times d}$ denote the behavioral signal vectors of the regression instances in the slice. We compute the centroid:
\begin{equation}
C_s = \operatorname{mean}\left(X_{0}[s]\right)
\label{eq:centroid}
\end{equation}
Next, we compute the Euclidean distance of each regression instance to the centroid and rank the distance from closest to furthest. Finally, we select the closest $K$ regression instances for execution on LLMUT-V1.

\paragraph{\texttt{Furthest point sampling (FPS)}}

Given a slice $s$, we first compute the centroid of the slice using Equation \ref{eq:centroid}.

We initialize the sampling process by selecting the regression instance closest to the centroid:

\begin{equation}
i_1 = \arg\min_{i \in s} \|x_i - C_s\|_2.
\end{equation}

Let $S$ denote the set of selected indices, initialized as $S = \{i_1\}$. For each remaining selection step, we compute the distance of each candidate regression instance to its nearest selected point:

\begin{equation}
d(i, S) = \min_{j \in S} \|x_i - x_j\|_2.
\end{equation}

The next point is selected using a greedy farthest-point criterion:

\begin{equation}
i_{t} = \arg\max_{i \in s \setminus S} d(i, S).
\end{equation}

The selected set is updated as $S \leftarrow S \cup \{i_t\}$ until $|S| = K$, where $K$ is the desired number of samples per slice.

\subsubsection{Execute representative regression instance(s) on LLMUT-V1}
At this stage, we execute the selected regression instances on LLMUT-V1 and collect the execution results and agent conversation logs. 
\subsubsection{Extract signals from the executed regression instance(s) at LLMUT-V1}
We parse the agent conversation logs and compute the behavioral signals as described in \ref{subsec:signal-ref}. We also compute the task performance metric, patch success in the software issue resolution domain, from the execution results. Next, we normalize the signals using the same scaler we used for the \texttt{V0 dataset}, obtaining $X_i^{(1)} \in \mathbb{R}^{K \times d}$, a matrix with $K$ rows and $d$ behavioral signals.

\subsubsection{Classify change impact}

To classify the impact of a change on individual slices, we formulate the problem as a supervised classification task. The classification model uses two features designed to capture both the observed behavioral change and the reliability of its estimation: the change in failure rate and the sampling ratio.

The first feature used by the classification model is the change in failure rate, $\Delta FR(s)$, which captures the observed behavioral difference between the two versions for slice $s$:

\[
\Delta FR(s)=FR_{V1}(s)-FR_{V0}(s)
\]

where $FR_{V0}(s)$ and $FR_{V1}(s)$ denote the failure rates of the executed regression instances in versions $V0$ and $V1$, respectively. For each slice $s$, each executed regression instance has a binary \texttt{patch\_success} outcome, where a value of 1 indicates a successful patch and a value of 0 indicates a failed patch. The failure rates are computed as:

\[
FR_{V0}(s)=
\frac{N_{\mathrm{fail},V0}(s)}
{K},
\qquad
FR_{V1}(s)=
\frac{N_{\mathrm{fail},V1}(s)}
{K}
\]

where $N_{\mathrm{fail},V0}(s)$ and $N_{\mathrm{fail},V1}(s)$ represent the number of executed regression instances with \texttt{patch\_success}=0 in versions $V0$ and $V1$, respectively, and $K$ is the number of regression instances sampled and executed per slice as defined earlier.

We select $\Delta FR(s)$ as a feature because it directly represents the observed impact signal used to identify changes in slice behavior. The sign of $\Delta FR(s)$ captures the direction of the impact, where positive values indicate an increase in failures after the change and negative values indicate a reduction in failures. The magnitude of $\Delta FR(s)$ reflects the strength of the observed behavioral change, providing the primary signal for distinguishing regression and improvement behaviors.

However, $\Delta FR(s)$ alone does not capture the uncertainty associated with its estimation. Since CASPER estimates the impact using a sampled subset of available regression instances, the reliability of the observed failure-rate difference depends on the proportion of instances used in the estimation. Therefore, the sampling ratio is introduced as a second feature:

\[
\mathrm{sampling\_ratio}(s)=
\frac{K}
{|s|}
\]

where $|s|$ is the slice size or the number of regression instances in slice $s$.

The sampling ratio provides information about the confidence of the estimated impact. For example, a large positive $\Delta FR(s)$ obtained using only a small fraction of available regression instances may be caused by sampling variability, whereas the same observed change estimated using a larger fraction of instances provides stronger evidence of a real regression. By incorporating the sampling ratio, the classifier can account for uncertainty in the estimated impact instead of relying only on the observed failure-rate difference.

Using these two features, we train two independent binary classifiers. The regression classifier predicts whether a slice belongs to the regression category, while the improvement classifier predicts whether a slice belongs to the improvement category. Both classifiers are trained using historical changes where the impact labels are available, enabling them to learn the relationship between observed failure-rate changes, sampling coverage, and slice impact.

For each slice, the regression and improvement classifiers independently produce probabilities: $P_R(s)$ and $P_I(s)$ representing the likelihood of regression and improvement, respectively. The final impact category is determined by jointly considering these probabilities. If neither classifier assigns a probability above the decision threshold, the slice is classified as not changed. If both classifiers indicate a possible impact, the category with the larger probability is selected:

\[
\mathrm{Impact}(s)=
\begin{cases}
\mathrm{NC}, & P_R(s)<0.5 \land P_I(s)<0.5\\
\mathrm{R}, & P_R(s)\geq0.5 \land P_R(s)>P_I(s)\\
\mathrm{I}, & P_I(s)\geq0.5 \land P_I(s)>P_R(s)
\end{cases}
\]

This probability-based decision strategy allows regression and improvement predictions to be considered independently and thus avoids the error propagation that would result from a sequential classification pipeline.

\subsubsection{Compute change in failure impact score}

The change in failure impact score is computed through the following steps.

\begin{enumerate}[leftmargin=*]

\item \textit{Score regression instances using the learned failure prediction model.}
We use the decision function of the learned failure prediction model to score each regression instance in $X_i^{(1)}$, yielding the corresponding failure impact score ($f_i^{(1)}$).

\item \textit{Compute the change in failure impact score.}
For each executed regression instance in V1, we compute the change in failure impact score ($\Delta f_i$) by subtracting the failure impact score of the regression instance in V0 from the failure impact score of the regression instance in V1:
\begin{equation}
\Delta f_i = |f_i^{(1)} - f_i^{(0)}|
\label{eq:change_in_failure_score}
\end{equation}

\item \textit{Aggregate the change in failure impact score.}
For each slice, the change in failure impact score of executed regression instances can be aggregated using different aggregation functions such as \texttt{mean}, \texttt{median}, and \texttt{max}. 

\end{enumerate}

\subsubsection{Rank slices}
The aggregated change in the failure impact score is used to rank the slices by regression impact. Specifically, we first associate each slice with its classified change impact category (\texttt{R}, \texttt{I}, or \texttt{NC}) and its aggregated failure impact score. We prioritize slices based on the change impact category in the order \texttt{regressed} (\texttt{R}), \texttt{improved} (\texttt{I}), and \texttt{not changed} (\texttt{NC}). Within each category, slices are ranked in descending order of their aggregated failure impact score, such that slices with larger estimated impacts appear earlier in the ranking.
\subsection{Output}

The approach outputs a ranked list of slices by regression impact. This ranking enables regression testing prioritization by focusing execution on instances from slices with higher expected regression impact.
\section{Evaluation}
\label{sec:eval}
In this section, we present an empirical study to evaluate \casper{}.
\subsection{Application domain}
We select the software issue resolution application domain for our case study. In this application domain, the LLMUT is tasked with generating a patch that resolves a software issue, given the problem statement as input and the associated software codebase as context. The LLMUT contains a reasoning model (LLM) at its core that has access to tools that help it explore its environment and perform changes. The reasoning model interacts with the environment via a tool call and decides on the next action based on the tool call's response. The reasoning model's task execution is guided by users using system prompts and user prompts. The reasoning model's thoughts, actions (tool calls) and their results are recorded for analysis.
\subsection{Regression testing dataset}

We select the \texttt{SWE-bench verified dataset} \cite{jimenez2024swebench} as the regression testing dataset, as it is a widely used benchmark to evaluate software issue resolution agents and is manually curated to make sure each issue in the dataset has a clear problem description, its corresponding ground truth patches are correct, and the problem is solvable with the provided context. This dataset contains 500 instances spanning 12 open-source Python projects \cite{jimenez2024swebench}. Each instance contains the issue ID, the problem statement, the repository name, the base commit where the issue occurred, the ground-truth patch, and other metadata.

\subsection{Subject system}

We selected the \texttt{Mini-swe-agent}\footnote{\url{https://github.com/SWE-agent/mini-swe-agent/tree/v1}} \cite{yang2024sweagent} as our subject system based on the following criteria:
\begin{itemize}
    \item [$\blacksquare$] The LLMUT, excluding the reasoning model, needs to be open-source to access and modify the code and the prompt.
    \item [$\blacksquare$] The LLMUT needs to support open-source reasoning models for full reproducibility of the experiments.
    \item [$\blacksquare$] The LLMUT needs to support logging conversations between the reasoning model and its environment, as our approach needs to process the conversation logs.
    \item [$\blacksquare$] The LLMUT needs to include an evaluation harness for our selected regression testing dataset.
\end{itemize}

\texttt{Mini-swe-agent} is a widely adopted version of \texttt{swe-agent} \cite{yang2024sweagent} that supports all LLM models using libraries such as Litellm. The reasoning model interacts with the environment using bash commands. The scaffold executes the bash commands generated by the reasoning model and returns the result to the model. \texttt{Mini-swe-agent} supports local deployment as well as cloud deployment, and also provides scripts to run evaluation on the \texttt{SWE-bench dataset}.

%We selected \texttt{Devstral-Small 2-24B-Instruct-2512}\footnote{\url{https://huggingface.co/mistralai/Devstral-Small-2-24B-Instruct-2512}} because it was a state-of-the-art reasoning model supported by \texttt{Mini-swe-agent}, while remaining sufficiently compact to be deployed within our available hardware resources. This model has 24 billion parameters and a 256K context window. \texttt{Mini-swe-agent} with this model was evaluated using  \texttt{SWE-bench verified dataset} and obtained $56.4\%$ issue resolution rate\footnote{\url{https://www.swebench.com/}}.  

\subsection{Evaluation dataset preparation}
\label{subsec:ds_prep}
Since our approach requires access to the conversation logs between the reasoning model and its environment, we collected these logs from two sources. The first source is the \texttt{SWE-bench leaderboard}\footnote{\url{https://www.swebench.com/leaderboard}}, where conversation logs and evaluation results are available for each submission. The submission details are maintained in a GitHub repository\footnote{\url{https://github.com/swe-bench/experiments}}. The second source consists of experiments in which we simulated changes to \texttt{Mini-swe-agent} and collected the corresponding conversation logs and evaluation results for models that we could host using the available GPU resources.

\subsubsection{Model change dataset}
We consider both same-family model changes and different-family model changes to simulate real-world scenarios that represent model updates within the same model family and transitions across different model families, respectively.
From the first source, we selected the submission that used \texttt{claude-4.5-sonnet} with a high reasoning level as the baseline and \texttt{claude-4.5-opus} with a high reasoning level as the changed version (V1) to evaluate our approach on same-family model changes. Both the baseline and V1 versions used \texttt{mini-swe-agent} version 2.0.0 as the LLMUT. In addition, we selected the submission that used \texttt{Kimi-k2-thinking} as the changed version (V1) and \texttt{Devstral-Small 2-24B-Instruct-2512} as the baseline to evaluate our approach on different-family model changes. Both the baseline and V1 versions used \texttt{mini-swe-agent} version 1.17.5 as the LLMUT. Since the baseline is not available on the leaderboard, we collected its conversation logs and evaluation results by running the \texttt{SWE-bench verified dataset} on the baseline model.

\subsubsection{Prompt change dataset}
We constructed two prompt-change datasets from the second source because the first source only includes executions with default prompts. To evaluate the impact of prompt modifications, we simulated custom prompt changes and generated the corresponding execution outcomes. The first dataset was prepared by applying a prompt change on \texttt{Mini-swe-agent} version 1.17.5 with \texttt{Devstral-Small 2-24B-Instruct-2512} as the reasoning model, while the second dataset was prepared by applying a prompt change on \texttt{Mini-swe-agent} version 2.0.0 with \texttt{Devstral-Small 2-24B-Instruct-2512} as the reasoning model. The first prompt change diff is available in Appendix \ref{app:prompt-change-one-diff} and the second prompt change diff is available in Appendix \ref{app:prompt-change-two-diff}. The prompts for both are sourced from the developers' revisions of \texttt{Mini-swe-agent}. For the first prompt change, the baseline commit id is \texttt{e2453091} and the V1 commit id is \texttt{22d33edf}. For the second prompt change dataset, the baseline version commit id is \texttt{3ec65451} and the V1 commit id is \texttt{89c6c5d1}. We collected the conversation logs and evaluation results for the baseline and changed versions by executing the \texttt{SWE-bench verified dataset} on both versions.

The primary modification of the first prompt change, outlined in Appendix \ref{app:prompt-change-one-diff}, is to the submission instructions, which broaden the submission condition from completing code changes to completing all work (including reading, editing, and testing), rename the required submission command from MINI\_SWE\_AGENT\_FINAL\_OUTPUT to COMPLETE\_TASK\_AND\_SUBMIT\_FINAL\_OUTPUT, and explicitly state that no further reading, editing, or testing is allowed after submission. Overall, the update refines the task completion and submission semantics.

The second prompt change, outlined in Appendix \ref{app:prompt-change-two-diff}, primarily changes the interaction protocol between the reasoning model and the environment by replacing the previous text-based command format with explicit tool calls. The system prompt was simplified by removing strict formatting requirements such as the mandatory THOUGHT section and the requirement to produce exactly one custom mswea\_bash\_command code block containing a single command. Instead, the updated prompt requires reasoning text together with at least one bash tool call, allowing greater flexibility in issuing commands across interactions. The instructions and examples were revised accordingly, including changing the submission command to use a standard bash code block.

Table \ref{tab:res-rate} shows the issue resolution rate of the original and V1 versions for each dataset. The resolution rate is computed as the ratio of the number of issues resolved to the total number of issues in the dataset. The results show that the V1 versions have a higher resolution rate than the baseline versions for all datasets except for the first prompt change. This indicates that the changes made in V1 have improved the LLMUT's performance in resolving issues across most datasets, but not for prompt change one.

\begin{table}[!t]
\centering
\caption{The issue resolution rate of the datasets on \texttt{SWE-bench verified} dataset}
\label{tab:res-rate}

\begin{tabular}{@{}lcc@{}}
\toprule
Dataset name             & baseline resolution rate & V1 resolution rate \\ \midrule
Model change             & 56.4\%                   & 63.4\%             \\
Same-family model change & 71.4\%                   & 76.8\%             \\
Prompt change one        & 56.4\%                   & 50.6\%             \\
Prompt change two        & 37.8\%                   & 62.4\%             \\ \bottomrule
\end{tabular}%

\end{table}

\subsubsection{Ground truth dataset preparation}
The ground truth dataset for evaluating the impact ranking approach is prepared using the following steps across all datasets.
\begin{enumerate}
    \item Generate the slices using \proposedsi{}.
    \item For each slice:
        \begin{enumerate}
            \item We compute the \texttt{Failure ratio} by dividing the number of regression instances with \texttt{patch success} of 0 (failed patch) by the slice size.
            \item we compute the \texttt{Failure ratio} for the V1 version
            \item we compute the \texttt{change in failure ratio} by subtracting the \texttt{Failure ratio} of the baseline version from the \texttt{Failure ratio} of the V1 version. We use the absolute value of the \texttt{change in failure ratio} as an impact score and rank the slices in decreasing order of this score.
        
    \item  We label a slice as \texttt{"regressed"} if its \texttt{change in failure ratio} is greater than 0, \texttt{"not-impacted"} if the \texttt{change in failure ratio} is equal to 0, and \texttt{"improved"} if the \texttt{change in failure ratio} is less than 0. For evaluation purposes, we group \texttt{"not-impacted"} and \texttt{"improved"} slices into a single \texttt{not-regressed} category.
\end{enumerate}
    
\end{enumerate}

Table \ref{tab:regressed_slices} reports the number of slices and regressed slices for each dataset before and after balancing. Because the datasets exhibit class imbalance, we balanced the classes by randomly sampling from the majority class so that the numbers of regressed and non-regressed slices were equal. In addition to mitigating the bias caused by imbalanced class distributions, this balancing strategy enables evaluation across multiple balanced subsets of slices, allowing us to account for the variability introduced by different slice compositions. This is particularly important when regressed slices dominate the dataset, as random ranking can otherwise achieve high performance simply due to their prevalence rather than their ability to effectively prioritize them.

To account for randomness in the sampling process, we repeated the downsampling procedure \texttt{30} times with replacement. Each downsampled dataset was then evaluated using both our approach and the baseline method, and the final results are reported as averages over the \texttt{30} runs. To minimize the effect of randomness of the LLM, we set the temperature of the LLM to 0.0 for all settings.

\begin{table}[!t]
\centering
\caption{Regressed slices for each dataset before and after resampling}
\label{tab:regressed_slices}

\begin{tabular}{@{}lcccc@{}}
\toprule
Dataset & Slices & Regressed slices & Slices after resampling & Regressed slices after resampling \\ \midrule
Model change             & 52 & 13 & 26 & 13 \\
Same family model change & 52 &  7& 14 & 7 \\
Prompt change one        & 52 & 28 & 48 & 24  \\
Prompt change two        & 60 & 12 & 24  & 12  \\ \bottomrule
\end{tabular}%

\end{table}

\subsection{Baseline impact ranking approach}
Since we were unable to identify an existing baseline for slice ranking based on regression impact, we implemented a simple random-ranking baseline that assigns slices a random order.

\subsection{Conversation log parser implementation}
The conversation log is a JSON file that contains the messages exchanged between the reasoning model and its environment, along with other metadata. The message is a list, with each element a dictionary containing the message content. The message content schema differs between \texttt{Mini-swe-agent} versions 1.17.5 and 2.0.0, so we implemented two parsers for each version. From the message content, we extract the reasoning text as well as the tool call or command details. The message also contains the tool call result, from which we extract the return code. One message corresponds to one turn in the conversation, and the total number of turns is equal to the size of the message list. 

Once we collect the reasoning text, tool call, or command, and the tool call or command result from each message, we use regular expressions and pattern matching to extract the command name and the corresponding arguments. \texttt{Mini-swe-agent} is instructed to use Linux commands to interact with the environment. We determine the agent state by matching the command name to a predefined list of commands associated with each state. The command names associated with each state are listed in Table \ref{tab:agent-command-patterns}. The tool call result contains the return code of the executed command, which we use to determine whether the command succeeded. A return code of 0 indicates success, while any other value indicates failure.  

To determine the \texttt{REPRODUCTION} state, we look for words in the thought that indicate the agent is trying to reproduce the issue. In particular, we look for the words associated with the \texttt{REPRODUCTION} state listed in Table \ref{tab:agent-command-patterns}. If any of these words are found in the thought, we classify the agent as being in the \texttt{REPRODUCTION} state. For the \texttt{NAVIGATION, PATCH, and VALIDATION} states, we extract the affected file paths from the command-line arguments.

\begin{table}[]
\caption{Mapping between agent states and associated command patterns used for software issue resolution.}
\label{tab:agent-command-patterns}

\begin{tabular}{@{}lp{11cm}@{}}
\toprule
\textbf{State} & \textbf{Associated Linux Commands / Patterns} \\ \midrule

VALIDATION &
pytest, tox, nosetests, unittest, python -m pytest, python3 -m pytest, python -m django, django test, pylint, flake8, mypy, ruff, sphinx-build, make test, make check, cmake --build, python -c, python3 -c, Python test scripts (test*.py, *\_test.py), Python verification scripts (verify*.py, check*.py, eval*.py, validate*.py), test, tests/, test\_, \_test.py, verify, check, eval, validate, test.sh, verify.sh, check.sh, validate.sh \\

\midrule
REPRODUCTION &
reproduce, reproduction, replicate, recreate, replicating, replication, recreating, reproducing \\

\midrule
SETUP &
pip install, apt-get, conda, export, poetry install, npm install, cd, pushd, popd, source, echo, exit, wget, whoami, date, bash, true\\

\midrule
NAVIGATION &
ls, find, locate, tree, du, df, which, whereis, pwd, cat, grep, head, tail, less, more, rg, wc, awk, sed -n, diff, git show, git log, git diff, git status \\
\midrule
PATCH &
in-place editing (sed -i, perl -pi), patch application (patch, git apply), git workspace mutation (git checkout, git restore, git reset, git stash, git add), file copy and overwrite (cp), file move/rename (mv), file creation (touch, mkdir), heredoc writes ($<<$ EOF, $<<$ EOT, $<<$ END), output redirection overwrite ($>$ file), append redirection ($>>$ file), tee-based writes (tee), echo redirection to file \\

\midrule
CLEANUP &
rm, unlink, find . -delete \\
\midrule
SUBMIT &
COMPLETE\_TASK\_AND\_SUBMIT\_FINAL\_OUTPUT\\

\bottomrule
\end{tabular}%

\end{table}

\subsection{\proposedsi{} hyperparameter tuning}
To enhance the performance of our method, we identified the optimal hyperparameter ranges and tuned the NSGA-II algorithm.

The search space included the following hyperparameters within  specific ranges:
\begin{itemize}
    \item \texttt{Population\_size} : [50, 150]
    \item \texttt{Crossover\_prob} (crossover probability): [0.5, 1.0]
    \item \texttt{Mutation\_prob} (mutation probability): [0.0, 0.5]
    \item \texttt{Min\_slice\_size} (minimum number of regression instances per slice): [5,10]
    \item \texttt{Max\_slice\_size} (maximum number of regression instances per slice): [10,20]
    \item \texttt{Max\_generations} (The number of generations before stopping optimization): [100,10000]
    \item \texttt{Max\_mutations} (The maximum number of mutations to perform at a time): [1,5]
\end{itemize}

We used the \textit{Optuna} library\cite{optuna_2019}, employing its NSGA-II implementation to perform a multi-objective search over the defined parameter space. We conducted an Optuna hyperparameter search with 30 trials, tracking the hyperparameter values and corresponding objective scores for each trial.

Table \ref{tab:opt-hyperpm-best-parms} presents the selected hyperparameter configurations obtained separately for each evaluation dataset. For each dataset, we perform an independent hyperparameter search using \textit{Optuna}. 

To select a single configuration, we first identify the Pareto-optimal set over all evaluated hyperparameter settings with respect to the objective functions. From this non-dominated set, we select the knee-point solution as a representative trade-off configuration, corresponding to the region of maximum curvature in the Pareto front and balancing competing objectives.

\begin{table}[t]
\centering
\caption{Optimal hyperparameter configurations selected for the NSGA-II-based slice identification algorithm on each evaluation dataset.}
\label{tab:opt-hyperpm-best-parms}
\renewcommand{\arraystretch}{1.2}
\begin{tabular}{lccc}
\toprule
\textbf{Hyperparameter} & \textbf{model change and prompt change one} & \textbf{same-family model change} & \textbf{prompt change two} \\
\midrule
\texttt{Population\_size}   & 136 & 93 & 96 \\
\texttt{Crossover\_prob}    & 0.994 & 0.894 & 0.823 \\
\texttt{Mutation\_prob}     &  0.219& 0.375 & 0.468 \\
\texttt{Min\_slice\_size}   & 5 & 5 & 5 \\
\texttt{Max\_slice\_size}   & 19 & 16 &  17\\
\texttt{Max\_generations}   & 6907 & 5710 & 9423 \\
\texttt{Max\_mutations}     & 2 & 1 & 1 \\
\bottomrule
\end{tabular}
\end{table}

\subsection{Slice identification baseline approaches}
Clustering algorithms serve as a natural baseline for slice identification because they optimize a primary objective, such as coherence, while potentially improving secondary objectives as a side effect. Compared to search-based techniques, clustering approaches are generally faster, making them computationally efficient alternatives. Moreover, if clustering can produce coherent, meaningful slices, it provides a viable method for identifying them. For our evaluation, we selected Gaussian Mixture Models (GMM)\cite{bishop2006pattern} and Hierarchical Density-Based Spatial Clustering of Applications with Noise (HDBSCAN)\cite{mcinnes2017hdbscan} as the baseline approaches. These methods were chosen because they represent two complementary families of clustering techniques, probabilistic and density-based, allowing us to assess slice identification across different clustering paradigms. GMM models each cluster as a multivariate Gaussian while HDBSCAN identifies clusters of varying densities and automatically labels low-density points as noise. Both approaches are well-suited to our original dataset (V0), where behavioral signals are columns with continuous values.

\paragraph{Gaussian Mixture Models (\gmm)}

Gaussian Mixture Models (\gmm) are a generative, probabilistic approach to clustering that assumes the data is drawn from a finite mixture of multivariate Gaussian distributions. Each latent component corresponds to a distinct cluster with its own mean vector and covariance matrix, allowing the model to capture ellipsoidal cluster shapes. Unlike partition-based methods such as k-means, \gmm{} provides a "soft" assignment, in which each data point receives a posterior probability of belonging to each component. Final hard cluster assignments are typically determined by selecting the component with the highest responsibility for each point. Because \gmm{} operates directly on continuous vectors.

\paragraph{Hierarchical Density-Based Spatial Clustering of Applications with Noise (\hdbscan)}
Hierarchical Density-Based Spatial Clustering of Applications with Noise (\hdbscan)\cite{mcinnes2017hdbscan} is a density-based clustering algorithm that extends DBSCAN by constructing a hierarchical cluster tree from mutual reachability distances. Rather than requiring a fixed density threshold, \hdbscan{} extracts the most stable clusters from this hierarchy by optimizing cluster persistence. A key feature of the method is its ability to automatically determine the number of clusters and label low-density points as noise. Because \hdbscan{} is robust to irregular cluster shapes and can operate directly on high-dimensional embeddings, it is well-suited for capturing the inherent structure in our V0 dataset with respect to behavioral signals.

We also conducted hyperparameter optimization for the baseline methods, \gmm{} and \hdbscan{}, using single-objective optimization with the \textit{Optuna} framework. For \gmm{}, the optimization objective was the \textit{Silhouette score}, and the tuned hyperparameters were the \texttt{number of slices} and the \texttt{membership probability threshold}, with the number of slices explored in the range $[10, 50]$ and the probability threshold in $[0.5, 0.9]$. For \hdbscan{}, the optimization objective was also the \textit{Silhouette score}, with the tuned hyperparameter being the \texttt{minimum cluster size}, explored in the range $[5, 10]$. Each baseline was optimized over 30 Optuna trials. The optimal configuration for \gmm{} and \hdbscan{} corresponded to the highest \textit{Silhouette score} achieved across trials, ensuring meaningful slice identification. For \hdbscan{}, the optimal hyperparameters are \texttt{number of slices: 48, 26, and 46} for the model change and the first prompt change, the second prompt change, and the same-family model change, respectively. 

The optimal configuration for \gmm{}, the optimal configuration hyperparameter tuples (\texttt{the number of slices}, \texttt{membership probability threshold}) are \texttt{(48, 0.68), (11, 0.56), (44, 0.68)} for the model change and the first prompt change, the second prompt change, and the same-family model change,  respectively. 

\subsection{Slice identification evaluation metrics}

We define the following evaluation metrics to compare the proposed approach against the baseline methods and align with the desired properties of a slice. 

\paragraph{Output consistency (OC)}

For each identified slice, we compute the output consistency metric using Equation \ref{eq:oc_metric}, expressed as a percentage. OC ranges from 50\% (when half of the issues have a pass rate of 1(resolved)) to 100\% (when all issues in a slice have similar pass rates). Higher OC values indicate that the slice contains regression instances for which the LLMUT performs similarly.

\paragraph{Slice coherence metric (SCM)}

The slice coherence metric is defined as the average pairwise cosine similarity of the behavioral signal vectors of regression instances within a slice. Higher SCM values indicate more coherent slices in the behavioral signal space, meaning that the grouped issues exhibit more similar behavioral characteristics.

\subsection{Slice ranking evaluation metrics}

We use the metrics \texttt{Recall@K} and \texttt{Area Under the Recall@K Curve (AUC)} to evaluate the effectiveness of our slice ranking approach.

\texttt{Recall@K} measures the cost-effectiveness of an approach's ranking under a limited testing budget. Recall@K is computed as the percentage of regressed slices (true positives) retrieved when selecting the top $K$ slices ranked by \casper{}. It is calculated by dividing the number of regressed slices among the top $K$ ranked slices by the total number of regressed slices in the ground-truth dataset. In our evaluation, Recall@K values are computed on the balanced ground-truth datasets described above and averaged over 30 independent downsampling trials.

The \texttt{Area Under the Recall@K Curve (AUC)} summarizes the overall cost-effectiveness of an approach across different testing budgets. It is computed as the area under the Recall@K curve, where the x-axis represents the percentage of slices inspected (Top-$K$\%) and the y-axis represents the corresponding Recall@K value. A higher AUC indicates that the approach consistently retrieves a larger proportion of regressed slices while requiring fewer slices to be tested across a range of testing budgets.

\subsection{\casper{} evaluation Experimental setup}

We evaluate variants of the \casper{} using a full-factorial experimental design with 12 configurations. The configurations are obtained by combining two failure prediction models, two representative regression instance sampling strategies, and three slice-level aggregation strategies.

For failure prediction, we evaluate Logistic Regression (LR) and Support Vector Machine (SVM) models, which achieved the best performance among the evaluated models in our preliminary experiments. We experimented with several classification models and selected LR and SVM as representative linear and non-linear approaches. For LR, we use the \texttt{liblinear} solver with a maximum of 1,000 iterations and an inverse regularization strength (\texttt{C}) of 1.0. For SVM, we use the radial basis function (RBF) kernel, as it consistently achieved the best performance across all evaluation datasets in preliminary experiments.

For change type classification, we evaluate five classification models: Logistic Regression (LR), Support Vector Machine (SVM), Decision Tree (DT), Random Forest (RF), and Gradient Boosting (GB). These models were selected to represent both linear and non-linear decision boundaries, as well as single-tree and ensemble learning approaches. Hyperparameters are kept at their default scikit-learn settings, except for fixing the random seed where applicable to ensure reproducibility.

Model selection is performed using leave-one-dataset-out (LODO) cross-validation to evaluate generalization across different software changes. In each fold, one regression testing dataset is held out for testing while the remaining regression testing datasets are used for training. Within the training partition, five-fold stratified cross-validation is used to compare candidate classification models and select the model with the highest mean ROC-AUC. The selected model is then retrained on the entire training partition and evaluated on the held-out regression testing dataset. This procedure is repeated such that each regression testing dataset serves as the held-out test set exactly once.

To identify representative regression instances for execution in \texttt{V1}, we evaluate two sampling strategies: \texttt{Closest Distance Sampling (CDS)} and \texttt{Furthest Point Sampling (FPS)}. We further evaluate different sample sizes, defined as the number of regression instances selected for execution from each slice at \texttt{V1}. The evaluated sample sizes range from one to the minimum slice size minus one across all datasets, corresponding to sample sizes from one to four. We exclude the minimum slice size itself to avoid the trivial case in which all samples from the smallest slices (i.e., slices containing 5 samples) are executed.

For aggregating changes in failure impact scores at the slice level, we evaluate three aggregation strategies: \texttt{mean}, \texttt{median}, and \texttt{max}.

\subsection{Research questions}
We aim to answer the following research questions in this evaluation case study.
\begin{itemize}[leftmargin=1cm]
    \item [RQ1:] How effective is our proposed slice identification approach?
    
    \textbf{Metrics:} Output Consistency (OC), and Slice Coherence Metric (SCM)
    \item [RQ2:] Which variant of the proposed slice-ranking approach is most effective at prioritizing regressed slices?

    \textbf{Metrics:} Recall@K and AUC
    \item [RQ3:] How effectively does the best slice-ranking variant prioritize regressed slices compared to a random ranking baseline?
    
    \textbf{Metrics:} Recall@K and AUC

\end{itemize}

\section{Results}
\label{sec:result}
In this section, we present the evaluation results and answer the stated research questions.

\subsection{RQ1: Slice identification evaluation }

To investigate RQ1, we applied both our proposed slice identification approach and the baseline clustering methods, each with its optimized hyperparameter configuration. We evaluated the approaches on the three \texttt{V0} datasets: (i) the model change and first prompt change datasets that have the same LLMUT baseline versions, (ii) the same-family model change dataset, and (iii) the second prompt change dataset. To reduce the impact of randomness, we repeated slice generation for both \casper{} and the baseline methods \texttt{30} times.

Table~\ref{tab:rq1-slice-size-distribution} reports the distribution of slice sizes produced by each slice identification method. Since all evaluation metrics are computed at the slice level, the reported slice counts correspond to the total number of slices generated across the 30 repetitions. Specifically, if a dataset is partitioned into $N$ slices in a single run, the reported count is $30 \times N$. For example, the second prompt change dataset is partitioned into 60 slices per run, resulting in a total of 1,800 slices. \proposedsi{} generally produces a larger number of slices because it explicitly constrains the slice size within a predefined range during optimization. This constraint controls slice granularity by avoiding excessively large slices that require more executions to obtain representative estimates of regression impact and are more difficult to interpret, while also avoiding excessively small slices that provide limited coverage of the underlying regression behavior. In contrast, the clustering-based baselines determine the number and size of clusters based on their underlying clustering objectives without explicit control over slice granularity. Overall, \hdbscan{} produces substantially larger slices than both \proposedsi{} and \gmm{} across all evaluation datasets.

\begin{table}[!t]
\centering
\caption{Distribution of slice sizes produced by the proposed slice identification approach (\proposedsi{}) and the baseline clustering methods (\gmm{} and \hdbscan) across the evaluation dataset slices.}
\label{tab:rq1-slice-size-distribution}
\renewcommand{\arraystretch}{1.2}

\begin{tabular}{llrrrrrr}
\toprule
\textbf{Dataset} &
\textbf{Method} &
\textbf{Count} &
\textbf{Mean} &
\textbf{Median} &
\textbf{Std} &
\textbf{Min} &
\textbf{Max} \\
\midrule

\multirow{3}{*}{Model change \& Prompt change one}
& \textbf{\proposedsi{}} & 1560 & 9.62 & 6.0 & 5.78 & 5 & 19 \\
& \gmm{}         & 1440 & 9.79 & 7.0 & 7.63 & 1 & 38 \\
& \hdbscan{}     & 360  & 32.42 & 21.5 & 22.69 & 12 & 85 \\
\midrule

\multirow{3}{*}{Prompt change two}
& \textbf{\proposedsi{}} & 1800 & 8.33 & 5.0 & 4.86 & 5 & 17 \\
& \gmm{}         & 330  & 45.27 & 42.0 & 12.65 & 28 & 69 \\
& \hdbscan{}     & 300  & 38.50 & 23.5 & 29.40 & 10 & 95 \\
\midrule

\multirow{3}{*}{Same-family model change}
& \textbf{\proposedsi{}} & 1560 & 9.62 & 8.0 & 4.25 & 5 & 16 \\
& \gmm{}         & 1320 & 10.84 & 6.0 & 12.75 & 3 & 70 \\
& \hdbscan{}     & 480  & 20.81 & 18.0 & 12.37 & 8 & 52 \\
\bottomrule
\end{tabular}

\end{table}

\begin{table}[!t]
\centering
\caption{Distribution of output consistency (OC) (\%) produced by the proposed slice identification approach (\proposedsi{}) and the baseline clustering methods (\gmm{} and \hdbscan) across the evaluation dataset slices.}
\label{tab:rq1-output-consistency}
\renewcommand{\arraystretch}{1.2}

\begin{tabular}{llrrrrrr}
\toprule
\textbf{Dataset} &
\textbf{Method} &
\textbf{Count} &
\textbf{Mean (\%)} &
\textbf{Median (\%)} &
\textbf{Std} &
\textbf{Min (\%)} &
\textbf{Max (\%)} \\
\midrule

\multirow{3}{*}{Model change \& prompt change one}
& \textbf{\proposedsi{}} & 1560 & \textbf{97.13} & \textbf{100.00} & \textbf{5.84} & \textbf{73.68} & \textbf{100.00} \\
& \gmm{}         & 1440 & 78.02 & 78.17 & 15.65 & 50.00 & 100.00 \\
& \hdbscan{}     & 360  & 79.09 & 79.13 & 7.69 & 64.29 & 90.20 \\
\midrule

\multirow{3}{*}{Prompt change two}
& \textbf{\proposedsi{}} & 1800 & \textbf{98.40} & \textbf{100.00} & \textbf{4.55} & \textbf{80.00} & \textbf{100.00} \\
& \gmm{}         & 330  & 84.16 & 88.10 & 12.83 & 57.14 & 100.00 \\
& \hdbscan{}     & 300  & 81.34 & 80.00 & 12.69 & 60.00 & 100.00 \\
\midrule

\multirow{3}{*}{Same-family model change}
& \textbf{\proposedsi{}} & 1560 & \textbf{97.97} & \textbf{100.00} & \textbf{4.39} & \textbf{81.25} & \textbf{100.00} \\
& \gmm{}         & 1320 & 74.48 & 75.00 & 14.08 & 50.00 & 100.00 \\
& \hdbscan{}     & 480  & 80.08 & 80.91 & 9.59 & 60.00 & 100.00 \\
\bottomrule
\end{tabular}

\end{table}
Table~\ref{tab:rq1-output-consistency} summarizes the distribution of output consistency $OC$ for the slices generated by each slice identification approach. Across all evaluation datasets, \proposedsi{} achieves the highest output consistency, with a mean $OC$ exceeding 97\%. In contrast, the baseline methods consistently produce substantially lower output consistency and perform similarly to one another. Furthermore, the minimum output consistency achieved by \proposedsi{} (73.68\%) is higher than that of both \gmm{} (50.00\%) and \hdbscan{} (60.00\%), demonstrating that \casper{} consistently generates slices that contain regression instances with consistent patch success outcomes.

Table~\ref{tab:rq1-scm} reports the distribution of the slice coherence metric (SCM). \proposedsi{} achieves the highest mean SCM for all evaluation datasets except the same-family model change dataset, where \gmm{} has higher SCM. This observation may be explained by differences in the underlying behavioral signal distribution across datasets. The behavioral signals in the same-family model change dataset may better satisfy the distributional assumptions of \gmm{}, allowing it to produce partitions with higher internal coherence. Although the baseline methods exhibit higher minimum SCM values, \proposedsi{} consistently achieves higher average coherence, indicating that it generally produces slices with better semantic coherence while exhibiting greater variability across slices.

Table~\ref{tab:rq1-scm} reports the distribution of the slice coherence metric (SCM). \proposedsi{} achieves the highest mean SCM for all evaluation datasets except the same-family model change dataset, where \gmm{} achieves a higher SCM. One possible explanation for this observation is that the behavioral signal distribution in this dataset may better align with the distributional assumptions of \gmm{}, allowing it to identify partitions with higher internal coherence. In contrast, \proposedsi{} jointly optimizes output consistency and slice coherence, which may result in a different partitioning strategy that balances both objectives. Although the baseline methods achieve higher minimum SCM values, \proposedsi{} consistently obtains higher average coherence across datasets, indicating that it generally identifies coherent slices while allowing greater variation among slices to capture diverse regression behaviors.

\begin{table}[!t]
\centering
\caption{Distribution of slice coherence metric (SCM) produced by the proposed slice identification approach (\proposedsi{}) and the baseline clustering methods (\gmm{} and \hdbscan{}) across the evaluation dataset slices.}
\label{tab:rq1-scm}
\renewcommand{\arraystretch}{1.2}

\begin{tabular}{llrrrrrr}
\toprule
\textbf{Dataset} &
\textbf{Method} &
\textbf{Count} &
\textbf{Mean} &
\textbf{Median} &
\textbf{Std} &
\textbf{Min} &
\textbf{Max} \\
\midrule

\multirow{3}{*}{Model change \& prompt change one}
& \textbf{\proposedsi{}} & 1560 & \textbf{0.6761} & \textbf{0.7174} & \textbf{0.2046} & 0.1771 & 0.9782 \\
& \gmm{}                     & 1440 & 0.6369 & 0.6438 & 0.1466 & \textbf{0.3912} & \textbf{1.0000} \\
& \hdbscan{}                 & 360  & 0.5432 & 0.5537 & 0.1174 & 0.3465 & 0.6885 \\
\midrule

\multirow{3}{*}{Prompt change two}
& \textbf{\proposedsi{}} & 1800 & \textbf{0.6815} & \textbf{0.7627} & \textbf{0.2056} & 0.2180 & \textbf{0.9625} \\
& \gmm{}                     & 330  & 0.5279 & 0.5214 & 0.0636 & 0.4431 & 0.6609 \\
& \hdbscan{}                 & 300  & 0.5508 & 0.5300 & 0.0503 & \textbf{0.4963} & 0.6396 \\
\midrule

\multirow{3}{*}{Same-family model change}
& \proposedsi{}           & 1560 & 0.5644 & \textbf{0.5950} & \textbf{0.2148} & 0.2041 & \textbf{0.9580} \\
& \textbf{\gmm{}}            & 1320 & \textbf{0.6041} & 0.6042 & 0.1665 & \textbf{0.3205} & 0.9357 \\
& \hdbscan{}                 & 480  & 0.5591 & 0.5876 & 0.1295 & 0.2987 & 0.7468 \\
\bottomrule
\end{tabular}

\end{table}

\begin{table}[!t]
\centering
\caption{Kruskal--Wallis statistical test results comparing the slice identification approaches across the evaluation datasets.}
\label{tab:rq1-statistical-test}
\renewcommand{\arraystretch}{1.2}

\begin{tabular}{llrrr}
\toprule
\textbf{Dataset} &
\textbf{Metric} &
\textbf{H Statistic} &
\textbf{$p$-value} &
\textbf{$\epsilon^2$} \\
\midrule

\multirow{2}{*}{Model change \& prompt change one}
&  $OC$ & 1544.05 & $<0.0001$ & 0.459 \\
&  $SCM$ & 203.43 & $<0.0001$ & 0.060 \\
\midrule

\multirow{2}{*}{Prompt change two}
&  $OC$ & 1347.02 & $<0.0001$ & 0.554 \\
&  $SCM$ & 330.41 & $<0.0001$ & 0.135 \\
\midrule

\multirow{2}{*}{Same-family model change}
&  $OC$ & 1925.93 & $<0.0001$ & 0.573 \\
&  $SCM$ & 39.03 & $<0.0001$ & 0.011 \\
\bottomrule
\end{tabular}

\end{table}

\begin{table}[!t]
\centering
\caption{Post-hoc Dunn's test results comparing \proposedsi{} with baseline slice identification approaches. Holm correction is applied for multiple comparisons. These results are based on the metric distributions reported in Tables~\ref{tab:rq1-output-consistency} and~\ref{tab:rq1-scm}; the Dunn's test indicates whether the difference between approaches is statistically significant. A dash indicates no statistically significant difference.}
\label{tab:rq1-dunn-test}
\renewcommand{\arraystretch}{1.2}

\begin{tabular}{lllll}
\toprule
\textbf{Dataset} &
\textbf{Metric} &
\textbf{Comparison} &
\textbf{$p$-value} &
\textbf{Winner} \\
\midrule

\multirow{4}{*}{Model change \& prompt change one}
& $OC$
& \proposedsi{} vs \gmm{}
& $<0.0001$
& \proposedsi{} \\
& $OC$
& \proposedsi{} vs \hdbscan{}
& $<0.0001$
& \proposedsi{} \\
\cmidrule(lr){2-5}
& $SCM$
& \proposedsi{} vs \gmm{}
& $<0.0001$
& \proposedsi{} \\
& $SCM$
& \proposedsi{} vs \hdbscan{}
& $<0.0001$
& \proposedsi{} \\

\midrule

\multirow{4}{*}{Prompt change two}
& $OC$
& \proposedsi{} vs \gmm{}
& $<0.0001$
& \proposedsi{} \\
& $OC$
& \proposedsi{} vs \hdbscan{}
& $<0.0001$
& \proposedsi{} \\
\cmidrule(lr){2-5}
& $SCM$
& \proposedsi{} vs \gmm{}
& $<0.0001$
& \proposedsi{} \\
& $SCM$
& \proposedsi{} vs \hdbscan{}
& $<0.0001$
& \proposedsi{} \\

\midrule

\multirow{4}{*}{Same-family model change}
& $OC$
& \proposedsi{} vs \gmm{}
& $<0.0001$
& \proposedsi{} \\
& $OC$
& \proposedsi{} vs \hdbscan{}
& $<0.0001$
& \proposedsi{} \\
\cmidrule(lr){2-5}
& $SCM$
& \proposedsi{} vs \gmm{}
& $<0.0001$
& \gmm{} \\
& $SCM$
& \proposedsi{} vs \hdbscan{}
& $0.1026$
& -- \\

\bottomrule
\end{tabular}

\end{table}

We conducted Kruskal--Wallis H tests to compare the RQ1 metrics across the slice identification methods. Table~\ref{tab:rq1-statistical-test} reports the H statistic, $p$-value, and omnibus effect size ($\epsilon^2$) for each metric. The $\epsilon^2$ effect size quantifies the magnitude of differences among all evaluated slice identification methods (\proposedsi{}, \gmm{}, and \hdbscan{}), measuring the proportion of variability in the ranked metric values attributable to differences among the methods. Larger $\epsilon^2$ values indicate stronger differences among the approaches. All metrics showed statistically significant differences across methods ($p < 0.0001$). The effect size for output consistency (OC) is large ($\epsilon^2$ between 0.46 and 0.57), indicating substantial differences among the approaches, with \proposedsi{} achieving the highest average OC across all datasets. In contrast, the effect size for the slice coherence metric (SCM) is smaller ($\epsilon^2$ ranging from 0.011 to 0.135), suggesting that the differences among approaches are less pronounced for coherence. Nevertheless, \proposedsi{} achieves the highest mean SCM in two of the three evaluation datasets and remains competitive with the baseline methods in the remaining dataset.

To identify whether the observed differences are specifically in favor of \proposedsi{}, we performed post-hoc Dunn's tests with Holm correction to compare \proposedsi{} against each baseline approach. Dunn's test compares pairs of methods using ranked observations, whereas the Holm correction adjusts the resulting p-values to account for multiple comparisons and control the family-wise error rate. As shown in Table~\ref{tab:rq1-dunn-test}, \proposedsi{} achieves statistically significant improvements over both \gmm{} and \hdbscan{} for the OC metric across all evaluation datasets ($p < 0.0001$), confirming that its higher output consistency is statistically significant. For the SCM metric, the results are more nuanced. \proposedsi{} achieves statistically significant improvements over both baselines on the model change \& prompt change one, and prompt change two datasets, where it also obtains higher mean SCM values. However, for the same-family model change dataset, \gmm{} achieves a higher mean SCM than \proposedsi{} (0.6041 versus 0.5644), and the difference between the two approaches is statistically significant ($p < 0.0001$). In contrast, the difference between \proposedsi{} and \hdbscan{} is not statistically significant for this dataset ($p = 0.1026$). Overall, these results demonstrate that \proposedsi{} provides consistent improvements in output consistency while achieving competitive slice coherence across different types of LLM changes.

\begin{mdframed}[backgroundcolor=gray!10, linecolor=black, linewidth=1pt, roundcorner=5pt, innertopmargin=10pt, innerbottommargin=10pt, innerleftmargin=10pt, innerrightmargin=10pt, skipabove=10pt]
\textbf{Summary of RQ1 Findings:}
\proposedsi{} achieves substantially higher output consistency across all evaluation datasets while maintaining comparable or slightly better slice coherence in most cases compared to the baseline clustering approaches.

\end{mdframed}

\subsection{RQ2: Effectiveness of proposed impacted slice ranking variants}

In this RQ, we evaluate the effectiveness of the impacted slice-ranking variants defined by our full-factorial experimental design. Each dataset is evaluated using 12 configurations obtained by combining two failure prediction models, two sampling strategies, and three aggregation strategies. We independently evaluate these configurations for each sample size, where each sample size corresponds to the number of regression instances selected for execution from each slice. As discussed in Section~\ref {subsec:ds_prep}, we repeated the down-sampling to account for class imbalance 30 times for each dataset. Hence, we report the average metrics over the 30 runs.
Table~\ref{tab:isr-variant-auc-distribution} summarizes the AUC distribution across the 12 configurations for each sample size setting.

Across the model change, prompt change one, and prompt change two datasets, the maximum observed AUC difference remains below 0.08, indicating that impacted slice ranking performance is largely robust to the selection of ranking variants for these change types. In contrast, the same-family model change dataset exhibits higher sensitivity to variant selection, with the largest AUC ranges observed for three and four regression instances per slice (0.1533 and 0.1493, respectively). As shown in Table~\ref{tab:isr-variant-extremes}, the best-performing configurations use SVM models with CDS sampling, while the worst-performing configurations use LR models with FPS sampling, suggesting that the prediction model and sampling strategy have a stronger influence in this challenging setting. This increased sensitivity may be attributed to the characteristics of same-family model changes, which contain fewer regressed slices (7 out of 52) and likely exhibit subtler behavioral differences, resulting in weaker and less separable regression signals. Under such low-signal conditions, the choice of modeling configuration can have a greater impact on extracting useful patterns, whereas the aggregation strategy appears to have a smaller effect.

To further analyze the factors contributing to the larger performance differences, we examined the best- and worst-performing variants for cases where the AUC range exceeded 0.10. Table~\ref{tab:isr-variant-extremes} shows that the observed performance differences are mainly associated with the choice of failure prediction model and sampling strategy. Specifically, SVM-based failure prediction combined with CDS sampling consistently achieves higher AUC values, whereas LR-based prediction combined with FPS sampling produces lower performance. Although the aggregation strategy varies across these configurations, its contribution appears less pronounced than the effects of the prediction model and sampling approach. These results suggest that variant sensitivity is mainly driven by the performance of the failure prediction model under different sampling conditions, rather than by the aggregation strategy.

\begin{table}[t]
\centering
\caption{AUC distribution across impacted slice ranking variants for different numbers of regression instances sampled per slice. Each row summarizes the performance of 12 variant combinations. The largest AUC range within each dataset is highlighted, indicating the highest sensitivity to variant selection.}
\label{tab:isr-variant-auc-distribution}
\renewcommand{\arraystretch}{1.2}

\begin{tabular}{llcccccc}
\toprule
\textbf{Dataset} & \textbf{\texttt{n}} & \textbf{Best Variant} &
\textbf{Max AUC} & \textbf{Min AUC} & \textbf{Mean AUC} &
\textbf{Std.} & \textbf{Range} \\
\midrule

\multirow{4}{*}{Model change}
& 1 & LR-CDS-median   & 0.4835 & 0.4246 & 0.4539 & 0.0246 & 0.0589 \\
& 2 & LR-CDS-max      & 0.5254 & 0.5003 & 0.5100 & 0.0089 & 0.0252 \\
& 3 & LR-CDS-mean     & 0.5803 & 0.5215 & 0.5486 & 0.0191 & 0.0588 \\
& 4 & LR-CDS-median   & 0.6112 & 0.5382 & 0.5717 & 0.0246 & \textbf{0.0730} \\

\midrule

\multirow{4}{*}{Prompt change two}
& 1 & LR-FPS-max      & 0.5575 & 0.5282 & 0.5439 & 0.0077 & 0.0293 \\
& 2 & LR-FPS-mean     & 0.5706 & 0.5117 & 0.5424 & 0.0274 & \textbf{0.0588} \\
& 3 & SVM-CDS-median  & 0.6092 & 0.5571 & 0.5817 & 0.0228 & 0.0522 \\
& 4 & LR-FPS-median   & 0.6126 & 0.5676 & 0.5905 & 0.0194 & 0.0449 \\

\midrule

\multirow{4}{*}{Prompt change one}
& 1 & LR-FPS-max      & 0.5416 & 0.5337 & 0.5372 & 0.0031 & 0.0079 \\
& 2 & SVM-FPS-max     & 0.5957 & 0.5724 & 0.5849 & 0.0073 & 0.0233 \\
& 3 & LR-CDS-mean     & 0.6044 & 0.5641 & 0.5867 & 0.0138 & 0.0403 \\
& 4 & LR-CDS-median   & 0.6148 & 0.5627 & 0.5844 & 0.0156 & \textbf{0.0520} \\

\midrule

\multirow{4}{*}{Same-family model change}
& 1 & SVM-CDS-mean    & 0.4865 & 0.4637 & 0.4748 & 0.0077 & 0.0229 \\
& 2 & SVM-FPS-mean    & 0.4808 & 0.4323 & 0.4587 & 0.0133 & 0.0486 \\
& 3 & SVM-CDS-median  & 0.5202 & 0.3669 & 0.4401 & 0.0613 & \textbf{0.1533} \\
& 4 & SVM-CDS-median  & 0.5977 & 0.4485 & 0.5289 & 0.0614 & 0.1493 \\

\bottomrule
\end{tabular}

\end{table}

\begin{table}[!t]
\centering
\caption{Comparison of the best- and worst-performing impacted slice ranking variants for cases where the AUC difference exceeds 0.10.}
\label{tab:isr-variant-extremes}
\renewcommand{\arraystretch}{1.2}

\begin{tabular}{p{2.2cm}lccccccccc}
\toprule
\multirow{2}{*}{\textbf{Dataset}} &
\multirow{2}{*}{\textbf{\texttt{n}}} &
\multicolumn{4}{c}{\textbf{Best-performing Variant}} &
\multicolumn{4}{c}{\textbf{Worst-performing Variant}} &
\multirow{2}{*}{\textbf{Range}} \\
\cmidrule(lr){3-6}
\cmidrule(lr){7-10}
&
&
\textbf{Model} &
\textbf{Sampling} &
\textbf{Agg.} &
\textbf{AUC} &
\textbf{Model} &
\textbf{Sampling} &
\textbf{Agg.} &
\textbf{AUC} &
\\
\midrule

Same-family model change &
3 &
SVM &
CDS &
Median &
0.5202 &
LR &
FPS &
Mean &
0.3669 &
\textbf{0.1533}
\\

Same-family model change &
4 &
SVM &
CDS &
Median &
0.5977 &
LR &
FPS &
Median &
0.4485 &
0.1493
\\

\bottomrule
\end{tabular}

\end{table}

\begin{table}[t]
\centering
\caption{Model selection using leave-one-dataset-out (LODO) cross-validation. The best and second-best classifiers were identified based on the mean out-of-fold (OOF) ROC-AUC. A paired DeLong test on the OOF predictions was used to determine whether the observed difference was statistically significant. $\Delta$AUC denotes the absolute difference between the cross-validation ROC-AUC values of the two models. Since Random Forest was statistically indistinguishable from the best-performing model in three of the four datasets for both regression and improvement prediction, and the remaining statistically significant differences exhibited negligible practical effect sizes ($\Delta$AUC $\leq$ 0.00033), Random Forest was selected as the unified classifier for all subsequent experiments.}
\label{tab:casper-model-selection}

\renewcommand{\arraystretch}{1.2}
\begin{tabular}{lp{2cm}ccccc}
\toprule
\textbf{Classifier Type} &
\textbf{Dataset} &
\textbf{Best} &
\textbf{Second} &
\textbf{$\Delta$AUC} &
\textbf{$p$-value} &
\textbf{Significant} \\
\midrule

\multirow{4}{*}{Regression}
& Prompt change one &
Random Forest &
Decision Tree &
0.00028 &
0.2509 &
No \\

& Model change &
Random Forest &
Decision Tree &
0.00009 &
0.4794 &
No \\

& Same-family model change &
Decision Tree &
Random Forest &
0.00033 &
0.0072 &
Yes \\

& Prompt change two &
Random Forest &
Decision Tree &
0.00011 &
0.4546 &
No \\

\midrule

\multirow{4}{*}{Improvement}
& Prompt change one &
Random Forest &
Gradient Boosting &
0.00022 &
0.8839 &
No \\

& Model change &
Random Forest &
Gradient Boosting &
0.00038 &
0.7271 &
No \\

& Same-family model change &
Decision Tree &
Random Forest &
0.00031 &
0.0471 &
Yes \\

& Prompt change two &
Random Forest &
Decision Tree &
0.00094 &
0.1019 &
No \\

\bottomrule
\end{tabular}

\end{table}

Tables~\ref{tab:casper-model-selection} and~\ref{tab:casper-classifier-heldout} report the model selection and evaluation results of the classifiers used by \casper{} to predict the impact type of a change on individual slices. Since \casper{} independently models regression and improvement behaviors, we train two binary classifiers: one for identifying regressed slices and another for identifying improved slices. Model selection is performed exclusively on the training portion of each leave-one-dataset-out split using out-of-fold (OOF) cross-validation. Specifically, for each split, three change datasets are used for training and model selection, while the remaining dataset is reserved as an unseen test set and is not used during model selection. The classifier with the highest OOF ROC-AUC is initially identified as the best candidate, and paired DeLong tests are used to assess whether its improvement over the second-best model is statistically significant. After model selection, the final classifier is retrained using the complete training data and evaluated once on the held-out dataset.

For regression prediction, Random Forest consistently provides the strongest and most stable performance across datasets. It achieves the highest OOF ROC-AUC for three out of four held-out scenarios, with CV ROC-AUC values ranging from 0.9156 to 0.9230. Although Decision Tree achieves a slightly higher OOF ROC-AUC for the same-family model change scenario (0.9313 compared with 0.9310 for Random Forest), the observed difference is extremely small ($\Delta$AUC=0.0003). Similarly, the differences between Random Forest and the second-best classifiers in the remaining datasets are negligible ($\Delta$AUC $<0.0003$), and none provide a practically meaningful advantage. Therefore, considering its consistently competitive performance, robustness across different change scenarios, and lower sensitivity to dataset-specific variations, we select Random Forest as the unified regression impact classifier. 

%On the held-out datasets, 

For improvement prediction, Random Forest again demonstrates the most consistent performance across datasets. It achieves the highest OOF ROC-AUC for three out of four scenarios, with CV ROC-AUC values ranging from 0.9341 to 0.9476. Decision Tree obtains a marginally higher score for the same-family model change dataset; however, the difference compared with Random Forest is minimal ($\Delta$AUC=0.0003), despite being statistically significant according to the paired DeLong test. This difference is unlikely to represent a meaningful performance improvement in practice. The remaining comparisons also show very small effect sizes ($\Delta$AUC $\leq 0.001$), supporting the use of a single robust classifier rather than dataset-specific models. Consequently, Random Forest is selected as the unified improvement impact classifier.

Table~\ref{tab:casper-classifier-heldout} presents the performance of the random forest regression and improvement classifiers on the held-out test set for all datasets. The Random Forest regression classifier achieves ROC-AUC values between 0.78 and 0.89. The lowest ROC-AUC (0.78) was registered on the same-family model change dataset. On the other hand, the final Random Forest improvement classifier achieves held-out ROC-AUC values ranging from 0.82 to 0.89, with the lowest value corresponding to the same-family model change dataset. The consistent difference in performance between the same-family model change and the other datasets for both regression and improvement classifiers could be explained by the difference in the baseline reasoning model used at $v0$. Specifically, the prompt change, prompt change two, and model change datasets share the same baseline reasoning model, allowing the classifiers to learn behavioral feature--outcome relationships that remain applicable across these datasets. In contrast, the same-family model change dataset uses a different baseline reasoning model, introducing a different behavioral regime that is not well represented in the training data. Consequently, the learned associations between behavioral signals and outcome labels may not transfer effectively, resulting in reduced generalization performance on the same-family model change dataset.

To further investigate this hypothesis, we performed a within-dataset control experiment, where the Random Forest classifiers were trained and evaluated using only samples from the same target dataset. This setting measures the performance when the training and testing data are generated from the same behavioral regime and allows us to quantify the generalization gap introduced by cross-dataset transfer. For regression prediction, the within-dataset ROC-AUC values were 0.95, 0.96, 0.95, and 0.90 for model change, prompt change two, prompt change one, and same-family model change, respectively. Compared with the corresponding held-out evaluations, these results correspond to performance drops of 0.12, 0.07, 0.08, and 0.12 ROC-AUC, respectively. Similarly, for improvement prediction, the within-dataset ROC-AUC values were 0.97, 0.96, 0.97, and 0.98, resulting in performance drops of 0.10, 0.10, 0.08, and 0.15 ROC-AUC, respectively. Notably, the same-family model change dataset exhibits the largest generalization gap for both regression and improvement prediction. These results indicate that the reduced held-out performance on the same-family model change dataset could be due to the mismatch between the behavioral patterns learned from datasets using the original baseline reasoning model and those induced by the different baseline reasoning model used at $v0$ in the same-family setting.

Overall, the results in Table~\ref{tab:casper-classifier-heldout} demonstrate that \casper{} can learn meaningful change impact patterns using the proposed classification framework. The Random Forest classifiers achieve promising ROC-AUC values across datasets, although performance varies depending on the change type and dataset characteristics. 
Since \casper{} uses the predicted change categories to construct the final ranking by prioritizing regression and improvement slices, accurate classification is essential for effective prioritization. 
Classification errors can place impacted slices in lower-priority categories or incorrectly prioritize non-impacted slices, thereby reducing the quality of the final ranking, both for regression and improvement. While there remains room for improvement, the obtained classification accuracy suggests that the learned signals are sufficiently informative to support downstream slice ranking. The effectiveness of the resulting ranking strategy is evaluated in the next research question.

\begin{table}[t]
\centering
\caption{Held-out evaluation performance of the Random Forest classifier for regression and improvement impact prediction. Following the model selection analysis, Random Forest is adopted across all datasets to avoid scenario-specific classifier selection and provide a consistent classification pipeline.}
\label{tab:casper-classifier-heldout}
\renewcommand{\arraystretch}{1.2}
\begin{tabular}{llccccc}
\toprule
\textbf{Classifier Type} &
\textbf{Dataset} &
\textbf{Model} &
\textbf{ROC-AUC} &
\textbf{Precision} &
\textbf{Recall} &
\textbf{F1} \\
\midrule

\multirow{4}{*}{Regression}
& Prompt change one &
\multirow{4}{*}{Random Forest} &
0.8721 &
0.7900 &
0.8819 &
0.8335 \\

& Model change &
&
0.8309 &
0.6975 &
0.7981 &
0.7444 \\

& Same-family model change &
&
0.7779 &
0.6938 &
0.6607 &
0.6768 \\

& Prompt change two &
&
0.8917 &
0.8471 &
0.7500 &
0.7956 \\

\midrule

\multirow{4}{*}{Improvement}
& Prompt change one &
\multirow{4}{*}{Random Forest} &
0.8924 &
0.7231 &
0.8393 &
0.7769 \\

& Model change &
&
0.8717 &
0.6909 &
0.7957 &
0.7396 \\

& Same-family model change &
&
0.8251 &
0.5664 &
0.7364 &
0.6403 \\

& Prompt change two &
&
0.8609 &
0.8435 &
0.8033 &
0.8229 \\

\bottomrule
\end{tabular}
\end{table}

\begin{mdframed}[backgroundcolor=gray!10, linecolor=black, linewidth=1pt, roundcorner=5pt, innertopmargin=10pt, innerbottommargin=10pt, innerleftmargin=10pt, innerrightmargin=10pt, skipabove=10pt]
\textbf{Summary of RQ2 Findings:}
\casper{} is largely robust to the choice of failure prediction model, sampling strategy, and aggregation method. Across most datasets, no single variant consistently outperforms the others, resulting in only modest differences in ranking performance. However, for same-family model changes, the choice of variant has a greater influence on ranking performance, with the prediction model choice and sampling strategy emerging as the primary differentiating factor. In addition, the regression and improvement classifiers used by \casper{} achieve strong predictive performance under held-out evaluation. The results show that a Random Forest classifier provides imperfect but promising accuracy, with variations across datasets. Though there is room for improvement, such classification models can help drive slice ranking, as discussed in the next research question.
\end{mdframed}

\subsection{RQ3: Effectiveness of proposed slice ranking approach against baseline}

\begin{figure*}[tb]
\centering

\begin{subfigure}{0.48\linewidth}
    \centering
    \includegraphics[width=\linewidth]{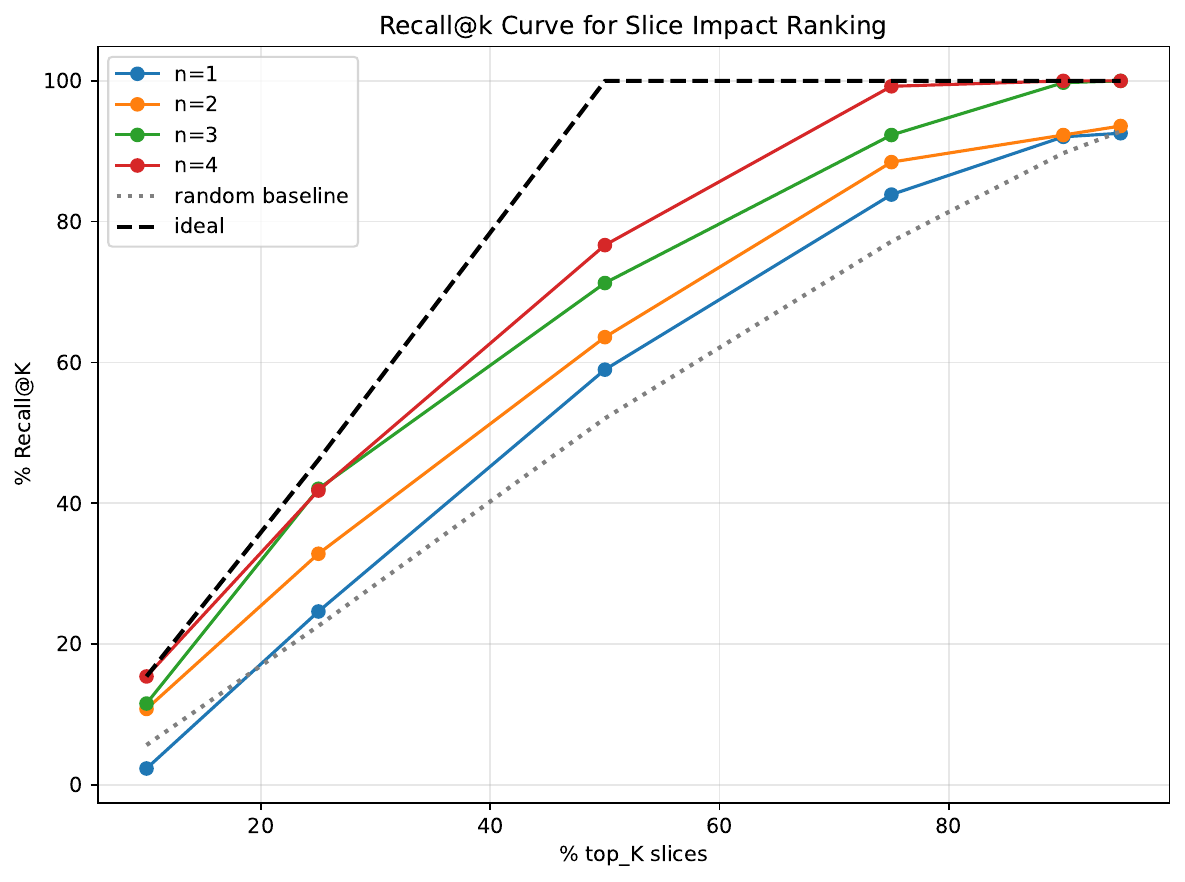}
    \caption{Model change, LR-CDS-mean variant.}
    \label{fig:model_change}
\end{subfigure}
\hfill
\begin{subfigure}{0.48\linewidth}
    \centering
    \includegraphics[width=\linewidth]{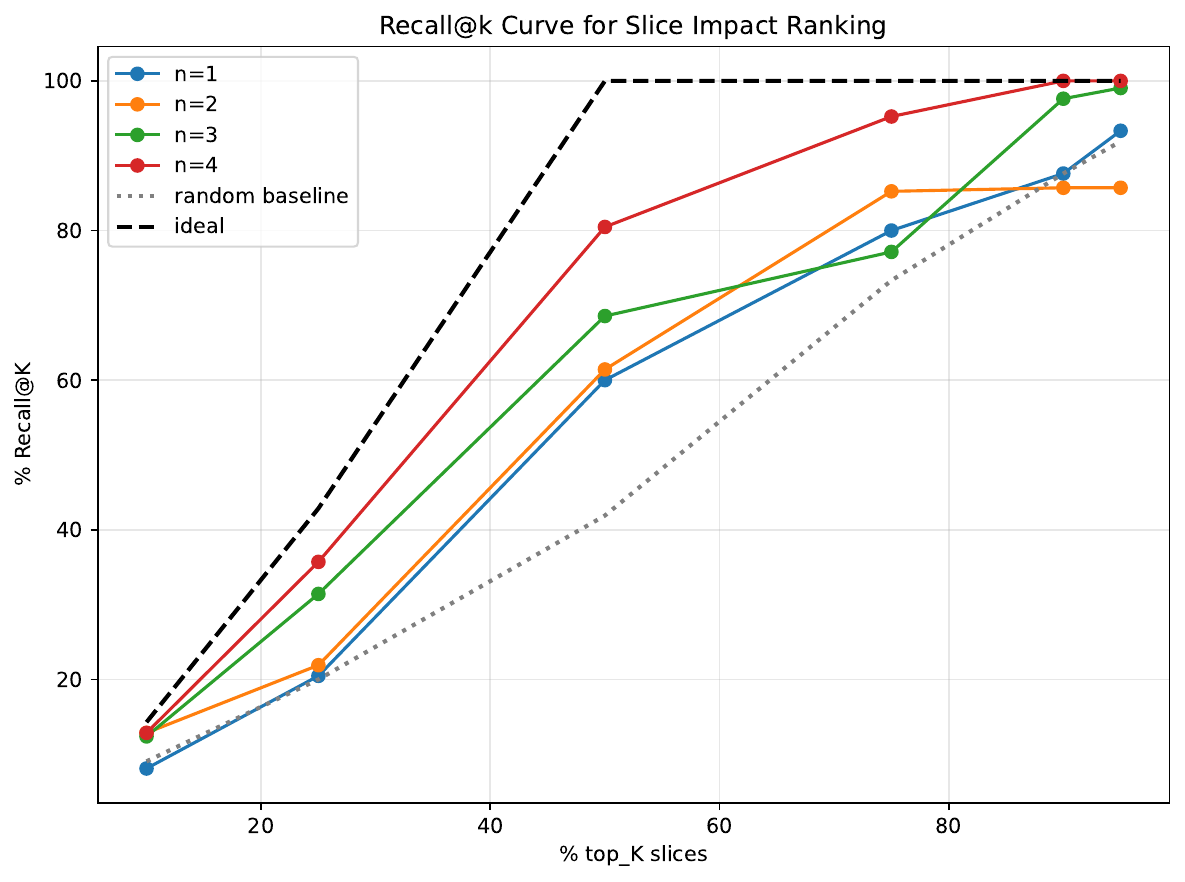}
    \caption{Same-family model change, SVM-CDS-median variant.}
    \label{fig:same_family}
\end{subfigure}

\vspace{0.5em}

\begin{subfigure}{0.48\linewidth}
    \centering
    \includegraphics[width=\linewidth]{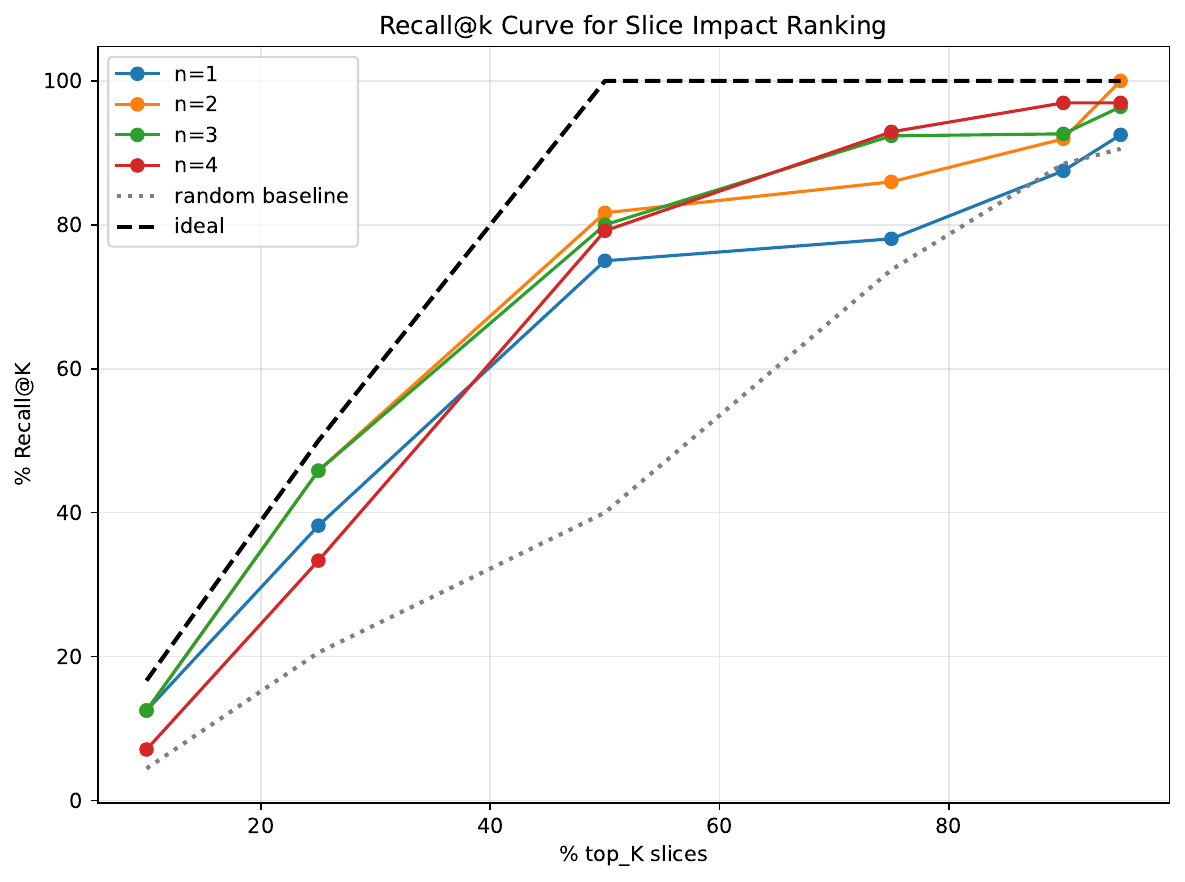}
    \caption{Prompt change 1, LR-CDS-max variant.}
    \label{fig:prompt_change}
\end{subfigure}
\hfill
\begin{subfigure}{0.48\linewidth}
    \centering
    \includegraphics[width=\linewidth]{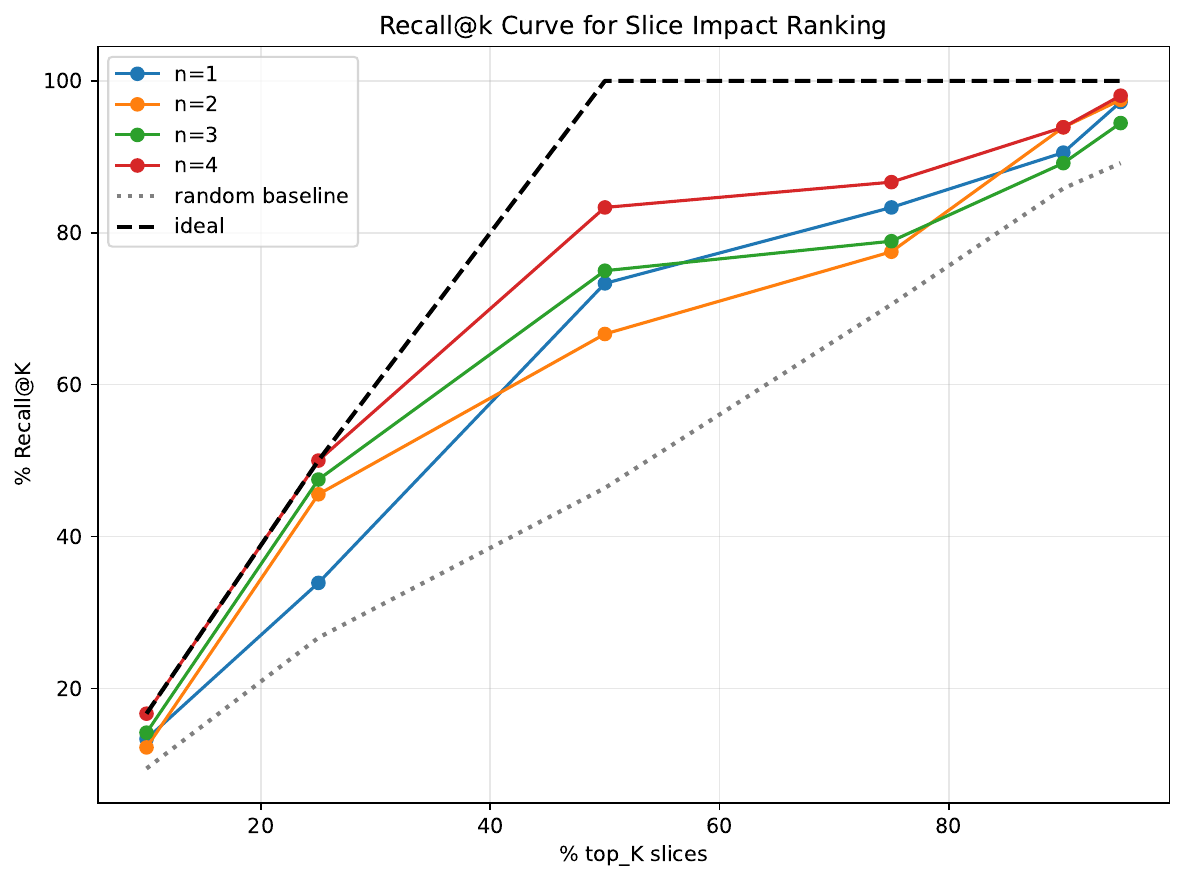}
    \caption{Prompt change 2, LR-FPS-median variant.}
    \label{fig:prompt_change2}
\end{subfigure}

\caption{Recall@K curves for \casper{} and the baseline under different numbers of samples per slice. The x-axis is the top-$k$ percentage obtained by dividing $K$ by the total number of slices in the ground-truth dataset. \texttt{n} denotes the number of samples per slice.}
\Description{Four line plots comparing Recall@K curves of \casper{} and baseline across different top-k percentages. The plots correspond to model change, same-family model change, prompt change one, and prompt change two datasets. Each plot shows recall performance as the number of samples per slice varies. The curves demonstrate differences among ranking approaches, with higher curves indicating greater effectiveness in slice ranking.}

\label{fig:rq3}
\end{figure*}

In this research question, we evaluate the effectiveness of our proposed impacted slice ranking approach using the Recall@K metric and compare it against a random slice ranking baseline and the ideal ranking. Figure~\ref{fig:rq3} presents the Recall@K curves for all evaluation datasets.
We repeated the evaluation 30 times to account for random variations during resampling for class balance (discussed in Section~\ref {subsec:ds_prep}). Hence, the reported Recall@K values are averages over the 30 runs for both \casper{} and the random ranking baseline.
For each dataset, we report the variant that achieved the highest AUC across all evaluated sample sizes ($n$).

For model changes, Figures~\ref{fig:model_change} and~\ref{fig:same_family} show that the Recall@K curves of \casper{} are consistently closer to the ideal Recall@K curve than the baseline, particularly as the number of regression instances sampled per slice ($n$) increases. \casper{} also achieves slightly better performance for the model change dataset than for the same-family model change. To investigate this difference, we quantified behavioral divergence between versions by computing the Mahalanobis distance between the standardized behavioral signal vectors before and after each change. Signal features were first standardized using the baseline behavioral distribution, after which change vectors were computed for every instance. A single covariance matrix was then estimated from the combined change vectors across all change types, allowing behavioral divergence to be measured in a common feature space. Comparing the resulting distance distributions showed that model changes exhibit significantly greater behavioral divergence than same-family model changes (Mann--Whitney $U=97{,}092$, $p<0.001$), with a higher median Mahalanobis distance (2.601 vs.\ 2.325). Although the effect size is modest (Cliff's $\delta=0.223$), the difference is consistent across the benchmark and indicates that transitioning between different model families (Devstral to Kimi K2) generally induces larger behavioral shifts than updating within the same model family (Claude 4.5 Sonnet to Claude 4.5 Opus). Consequently, the resulting changes in slice-level behavioral features are more pronounced, making impacted slices easier for \casper{} to distinguish and prioritize.

For prompt changes, the performance of \casper{} varies across the two prompt-change datasets, as shown in Figures~\ref{fig:prompt_change} and~\ref{fig:prompt_change2}. \casper{} achieves a Recall@K curve that more closely follows the ideal recall curve for the first prompt change than for the second. This observation is consistent with the measured behavioral divergence between the two prompt modifications. The first prompt change exhibits significantly greater behavioral divergence than the second (Mann--Whitney $U=156{,}590$, $p<0.001$), with a higher median Mahalanobis distance (3.305 vs.\ 2.936). Although the effect size is modest (Cliff's $\delta=0.253$), the difference is consistent across the benchmark and indicates that the first prompt modification induces larger behavioral changes overall. This larger behavioral divergence is also reflected in the broader impact of the first prompt modification, which introduces regressions across a substantially larger proportion of slices (28 of 52 compared with 12 of 60). Together, these results suggest that prompt modifications inducing larger behavioral shifts produce more pronounced changes in the slice-level behavioral features used by \casper{}, make impacted slices easier to distinguish and prioritize.

\begin{table}[t]
\centering
\caption{Comparison of the proposed impacted slice ranking approach and the random slice ranking baseline in terms of AUC computed from the averaged Recall@K curve over independent trials. Relative gain is computed with respect to the baseline AUC.}
\label{tab:rq3-auc-comparison}
\renewcommand{\arraystretch}{1.2}

\begin{tabular}{llccc}
\toprule
\textbf{Dataset} &
\textbf{\texttt{n}} &
\textbf{\casper{} AUC} &
\textbf{Baseline AUC} &
\textbf{Relative Gain (\%)} \\
\midrule

\multirow{4}{*}{Model change}
& 1 & 0.4813 & 0.4468 & 7.72 \\
& 2 & 0.5253 & 0.4468 & 17.58 \\
& 3 & 0.5803 & 0.4468 & 29.89 \\
& 4 & \textbf{0.6103} & 0.4468 & \textbf{36.59} \\
\midrule

\multirow{4}{*}{Same-family model change}
& 1 & 0.4680 & 0.4088 & 14.47 \\
& 2 & 0.4846 & 0.4088 & 18.55 \\
& 3 & 0.5202 & 0.4088 & 27.26 \\
& 4 & \textbf{0.5977} & 0.4088 & \textbf{46.21} \\
\midrule

\multirow{4}{*}{Prompt change one}
& 1 & 0.5400 & 0.4031 & 33.98 \\
& 2 & 0.5941 & 0.4031 & 47.40 \\
& 3 & \textbf{0.6025} & 0.4031 & \textbf{49.48} \\
& 4 & 0.5769 & 0.4031 & 43.13 \\
\midrule

\multirow{4}{*}{Prompt change two}
& 1 & 0.5427 & 0.4256 & 27.49 \\
& 2 & 0.5402 & 0.4256 & 26.92 \\
& 3 & 0.5637 & 0.4256 & 32.43 \\
& 4 & \textbf{0.6126} & 0.4256 & \textbf{43.92} \\
\bottomrule
\end{tabular}

\end{table}

Table~\ref{tab:rq3-auc-comparison} reports the relative improvement in AUC achieved by \casper{} over the random slice ranking baseline for different numbers of samples per slice ($n$) across all datasets. \casper{} consistently outperforms the baseline across all datasets and sampling configurations, including cases where only a single regression instance per slice is executed. The largest improvements are observed for the prompt change datasets, where the relative gains reach 49.48\% and 43.92\% for prompt change one and prompt change two, respectively. In contrast, the model change dataset shows more modest improvements, with gains ranging from 7.72\% to 46.21\%. Overall, increasing the number of samples per slice generally improves the ranking performance, as reflected by higher AUC values and larger relative gains for larger $n$, although the relationship is not strictly monotonic for all datasets.

\casper{} is particularly beneficial under limited regression testing budgets, as commonly encountered in CI/CD pipelines where frequent code changes make exhaustive regression analysis impractical. The cost of regression testing is influenced by both the amount of execution evidence required to identify changes and the effort required to analyze affected slices. Therefore, an effective prioritization approach should accurately identify regressed slices using limited execution evidence and rank them ahead of non-regressed slices so that developer attention can be focused on the most relevant changes.

As shown in Figure~\ref{fig:rq3}, \casper{} consistently achieves high recall at low top-$k$ budgets across different change scenarios, demonstrating its ability to place regressed slices among the highest-priority candidates. Since \casper{} first classifies slices into regression, improvement, and unchanged categories and then ranks slices according to their predicted impact within each category, the resulting ranking prioritizes regressed slices before improved and unchanged slices. Consequently, developers can analyze only a small subset of top-ranked slices while still discovering a substantial portion of the actual regressions, allowing limited debugging and analysis resources to be directed toward slices that require corrective action. Moreover, \casper{} maintains effective prioritization performance when fewer regression instances per slice ($n$) are available, demonstrating robustness when only limited execution evidence from the changed version can be collected. These results indicate that \casper{} enables efficient regression analysis by reducing the amount of evidence needed for prioritization and ensuring that limited developer effort is focused on slices with regression impact.

\begin{mdframed}[backgroundcolor=gray!10, linecolor=black, linewidth=1pt, roundcorner=5pt, innertopmargin=10pt, innerbottommargin=10pt, innerleftmargin=10pt, innerrightmargin=10pt, skipabove=10pt]
\textbf{Summary of RQ3 Findings:}
The proposed impacted slice ranking approach consistently outperforms random ranking, achieving higher Recall@K and AUC across all evaluated datasets and sample sizes. The improvement is most pronounced for prompt change datasets, while it is smaller for same-family model changes due to less pronounced distribution shifts between versions \texttt{V0} and \texttt{V1}. Furthermore, the approach is particularly beneficial under limited testing budgets, as commonly encountered in CI/CD pipelines with frequent code changes, by identifying regressed slices using limited execution evidence and prioritizing them ahead of improved and unchanged slices, allowing developers to focus their analysis efforts on regressed slices.
\end{mdframed}

\section{Threats to validity}
\label{sec:threats}
In this section, we discuss the threats to the validity of our study and the mitigation measures we implemented. 

\textbf{Internal threats to validity:} To reduce internal validity threats related to hyperparameter tuning for both \proposedsi{} and the baselines, we applied the same automated tuning procedure using Optuna with an identical search budget (30 trials). We also report the full tuning setup and results in the evaluation section. It is important to note that \proposedsi{} may require retuning when applied to different application domains or regression datasets. The impacted slice ranking evaluation relies on randomly generated balanced datasets obtained through downsampling of the majority class. Because different random samples may lead to slight variations in the evaluation metrics, we repeated the balancing procedure 30 times and report the average performance across all runs. This reduces the influence of sampling randomness and improves the robustness of the reported results.

\textbf{Construct threats to validity} Construct validity may be threatened by our use of the behavioral signals as a proxy for regression impact detection. The behavioral signals used in this study are engineered features intended to capture characteristics of the agent's reasoning and execution process. Although these signals were designed based on prior work and empirical observations, they may not fully represent all aspects of agent behavior that influence patch success. Consequently, some relevant behavioral factors may be omitted or only partially captured, potentially affecting the predictive performance of the failure prediction models used in the proposed impacted slice ranking approach.

\textbf{External validity threats:} We evaluated our approach using a single application domain, software issue resolution, which may limit the generalizability of our findings. However, this domain is a very active and important application domain for LLMs in software engineering. Further, note that our approach is not intended to be applied directly to any domain without adaptation. Applying it to a new domain requires customizing the behavioral signals and retuning the hyperparameters.

\textbf{Threats to conclusion validity:} We used several statistical tests when evaluating our approach, which may introduce threats to conclusion validity. To mitigate this, we repeated the experiments and measurements multiple times to reduce the impact of randomness and employed nonparametric statistical tests that do not rely on assumptions about the underlying data distribution.

\section{Related work}

\label{sec:related}
In this section, we discuss related work on regression testing for LLM-based systems and on slice identification techniques.
\subsection{Regression testing in LLM-based systems}
Regression testing is a well-established practice in traditional software systems to detect unintended changes in system behavior after code changes \cite{engstrom2010systematic,yoo2012regression}. However, the introduction of LLMs as reasoning engines in  LLM-based systems poses unique challenges that make conventional regression testing challenging. Ma \etal{} conducted a case study on toxicity detection and argued for re-examining the concept of regression testing for LLM-based systems due to the non-deterministic nature of LLMs, differing notions of correctness, and sensitivity to prompting ~\cite{ma2024my}. Unlike traditional software systems, where a failure in one regression instance indicates a regression, the authors argued that regression granularity should be slices, which are the sweet spot between tracking regression at the individual regression instance level and reporting performance drops across all regression instances, which is too coarse-grained. Slices should contain semantically related regression instances~\cite{johnson2023does} in which the LLMUT should have uniform performance, either uniformly performing poorly~\cite{eyuboglu2022domino, chung2019slice, johnson2023does} or uniformly successful, for regression instances in the slice to be useful for regression testing. This paper proposes the first practical, automated solution to further this objective.

\subsection{Slice identification techniques}
Several slice identification techniques from different application domains have been proposed~\cite{eyuboglu2022domino, zhang2022sliceteller, chung2019slice, olesen2024slicing, ghosh2025ladder, hua2022discover, plumb2022towards}.

Chung \etal~\cite{chung2019slice} proposed Slice Finder, an interactive framework that identifies interpretable data slices where a model performs poorly using two complementary techniques: decision-tree training, which yields naturally interpretable but non-overlapping slices by locating high-effect-size leaves and generalizing them upward, and lattice searching, which explores a broader space of potentially overlapping slices by traversing a discretized feature lattice with statistical significance testing and pruning.

Eyuboglu \etal~\cite{eyuboglu2022domino} proposed Domino, a slice identification and evaluation technique that performs slice identification in natural images, medical images, and time-series datasets. They leveraged cross-modal embeddings together with an error-aware mixture model to discover slices and generate slice descriptions by sourcing a corpus of candidate natural language phrases that identify slice characteristics, using cross-modal embeddings to semantically represent the slice, and then performing cosine-similarity-based matching against the candidate descriptions.

Building on the idea of identifying slices for model evaluation, the proposed slice identification method by Hua \etal~\cite{hua2022discover} specifically targets classification tasks. The approach, called \textbf{EDisa}, employs an error-distance-aware multivariate Gaussian mixture model. Unlike DOMINO~\cite{eyuboglu2022domino}, which combines Gaussian and categorical distributions, EDisa models all components as continuous variables: \(Z\) representing the embedding of each data point, \(E\) capturing the distance between the model’s prediction and the ground truth label, and \(y\) corresponding to the confidence score. To reduce dimensionality, the method applies PCA before clustering. However, EDisa has notable limitations: it is designed for encoder-only models and is restricted to classification datasets.

Zhang \etal~\cite{zhang2022sliceteller} proposed \textit{SliceTeller}, which incorporates a slice identification component that uses model predictions and metadata as input to perform slice search via frequent pattern mining techniques. Since this exhaustive search can generate a large number of slices, depending on the minimum support hyperparameter, the authors employ redundancy-pruning methods to reduce the search space while retaining meaningful slices.

Ghosh \etal~\cite{ghosh2025ladder} proposed LADDER slice identification, a method to identify and mitigate model errors by retrieving sentences that highlight attributes contributing to misclassifications. It projects the model’s internal image embeddings, the feature representations used by the classifier, into a visual-language space, and computes the difference between the mean embeddings of correctly classified and misclassified samples for each class. Top-K sentences whose text embeddings align with this difference are retrieved, and an LLM generates hypotheses describing attributes the model may be relying on incorrectly. Images that lack these attributes, as determined by similarity in the shared embedding space, form candidate error slices. Importantly, LADDER discovers and mitigates such errors \emph{without requiring human-provided attribute annotations}: it uses similarity scores to assign pseudo-labels indicating whether each attribute is present or absent. These pseudo-labels support two ensemble-based debiasing strategies, reweighting training samples (LADDER\textsubscript{reweight}) or constructing balanced datasets for fine-tuning (LADDER\textsubscript{bal}), with the latter demonstrating stronger empirical performance. This technique is limited to the image classification application domain and evaluated on natural and medical images.

The slice identification method proposed by Olesen\etal~\cite{olesen2024slicing} follows four main steps: it first derives image features directly from the model’s penultimate layer, eliminating the need for external embedding models; then it adds a supervised, low-dimensional fully connected layer trained with the same procedure as the original classifier to retain prediction-relevant information during dimensionality reduction. Using these learned embeddings, it applies Gaussian Mixture Models and BIC-based selection to cluster positive and negative samples separately, capturing coherent groups with similar error patterns. Finally, it evaluates and prioritizes these clusters using the Brier score, chosen for its threshold independence, calibration sensitivity, and comparability across cluster sizes, and identifies the strongest and weakest slices based on bootstrap-based uncertainty estimates.

While prior work has advanced slice identification across structured data, images, and classification tasks, existing methods generally remain limited in scope and rely on assumptions about input modality that do not translate to LLM-based systems that take textual input, which are more common. In contrast, our approach is not restricted to classification and naturally extends to generative tasks, the setting where LLMs are most impactful, allowing slice identification for outputs such as summaries, answers, or generated code. Unlike DOMINO, which depends on a predefined corpus of candidate slice descriptions, our method does not require any external topic inventory or manually curated phrase sets, enabling more flexible and adaptive slice characterization. Moreover, because it does not rely on image-specific architectures or encoder-only models as in EDisa or LADDER, our technique can be customized to virtually any domain by modifying the input dataset for the slice identification stage and the prompt components for the description generation stage. By removing assumptions tied to specific data modalities, tasks, or auxiliary resources, our approach broadens its applicability and complements prior techniques. This general, model-agnostic, and domain-flexible slicing enables the construction of slices that can serve as consistent units of regression when evaluating LLM-based systems.

\section{Conclusion}
\label{sec:conclusion}
Effective regression testing for LLM-based systems requires approaches that account for their unique characteristics. Unlike conventional software, where a test failure often indicates a defect, LLM-based systems may occasionally produce incorrect outputs due to their non-deterministic behavior, making individual test failures less informative. Moreover, evaluating performance only at the test suite level can be too coarse to identify which behaviors are affected by a system change. Therefore, regression testing requires mechanisms that organize regression instances into meaningful units and prioritize the units most likely to be regressed by changes.

To address this challenge, we propose \casper{}, a change-aware slice prioritization framework for regression testing of LLM-based applications. \casper{} identifies regression slices containing coherent regression instances with consistent performance characteristics and leverages behavioral signals extracted from agent execution logs to prioritize slices according to their likelihood of being impacted by a system change. By combining meaningful slice identification with change-aware prioritization, \casper{} enables more cost-effective regression testing under limited testing budgets.

We evaluated \casper{} in the software issue resolution domain using the SWE-bench Verified dataset, a highly active and important application domain for LLMs. The results demonstrate that the slice identification component achieves substantially higher output consistency while maintaining comparable or better slice coherence than clustering-based baselines. Furthermore, the regressed slice ranking approach consistently outperforms random ranking across different types of LLM-based system changes and testing budgets, demonstrating its effectiveness in prioritizing regressed slices during regression testing.

Although our evaluation focuses on software issue resolution, the proposed framework applies to a broader range of LLM-based systems. By adapting the behavioral signals and performance criteria to the target application domain, \casper{} provides a general approach for change-aware regression testing of evolving LLM-based systems.

Future work will investigate the applicability of \casper{} under a broader range of LLM-based system changes, including modifications to retrieval components, tool integrations, and inference configurations. Additionally, we plan to explore scalability improvements to enable the application of \casper{} to larger regression testing datasets, including more efficient slice identification and incremental updates as new regression instances or system changes are introduced.

\textbf{Replication Package}
%To facilitate replication and further research, we provide a replication package containing the implementation of \casper{}, the slice identification and regressed-slice ranking pipelines, configuration files, experimental scripts, and the data required to reproduce the reported results.\footnote{\url{https://anonymous.4open.science/r/casper_replication_package}}
To facilitate replication and further research, we will provide a replication package containing the implementation of \casper{}, the slice identification and regressed-slice ranking pipelines, configuration files, experimental scripts, and the data required to reproduce the reported results after accepetance for publication.

\section*{Acknowledgments}

This work was supported by a research grant from Huawei Canada, the Discovery Grant and Canada Research Chair programs of the Natural Sciences and Engineering Research Council of Canada (NSERC), and a Research Ireland grant 13/RC/2094-2.

\bibliographystyle{ACM-Reference-Format}
\bibliography{citation}

\appendix
\section{Appendix}
\subsection{Prompt Diff file for the prompt change dataset one}
\label{app:prompt-change-one-diff}

\begin{lstlisting}[
    language=Python,
    basicstyle=\scriptsize\ttfamily,
    breaklines=true,
    columns=fullflexible,
    keepspaces=true,
    frame=single
]

diff --git a/src/minisweagent/config/extra/swebench.yaml b/src/minisweagent/config/extra/swebench.yaml
index 37552510..15809dfa 100644
--- a/src/minisweagent/config/extra/swebench.yaml
+++ b/src/minisweagent/config/extra/swebench.yaml
<unchanged content>

## Submission

- When you've completed your changes or can't make further progress
+ When you've completed your work (reading, editing, testing), and cannot make further progress

  issue exactly the following command:

  ```bash
- echo MINI_SWE_AGENT_FINAL_OUTPUT && git add -A && git diff --cached
+ echo COMPLETE_TASK_AND_SUBMIT_FINAL_OUTPUT && git add -A && git diff --cached
\end{lstlisting}

\subsection{Prompt Diff file for the prompt change dataset two}
\label{app:prompt-change-two-diff}

\begin{lstlisting}[
    language=Python,
    basicstyle=\scriptsize\ttfamily,
    breaklines=true,
    columns=fullflexible,
    keepspaces=true,
    frame=single
]
diff --git a/src/minisweagent/config/benchmarks/swebench.yaml b/src/minisweagent/config/benchmarks/swebench.yaml
index c2ccfadf..d473af1d 100644
--- a/src/minisweagent/config/benchmarks/swebench.yaml
+++ b/src/minisweagent/config/benchmarks/swebench.yaml
@@ -1,20 +1,6 @@
 agent:
   system_template: |
-    You are a helpful assistant that can interact multiple times with a computer shell to solve programming tasks.
-    Your response must contain exactly ONE bash code block with ONE command (or commands connected with && or ||).
-
-    Include a THOUGHT section before your command where you explain your reasoning process.
-    Format your response as shown in <format_example>.
-
-    <format_example>
-    THOUGHT: Your reasoning and analysis here
-
-    ```mswea_bash_command
-    your_command_here
-    ```
-    </format_example>
-
-    Failure to follow these rules will cause your response to be rejected.
+    You are a helpful assistant that can interact with a computer shell to solve programming tasks.
   instance_template: |
     <pr_description>
     Consider the following PR description:
@@ -29,8 +15,7 @@ agent:
     You're a software engineer interacting continuously with a computer by submitting commands.
     You'll be helping implement necessary changes to meet requirements in the PR description.
     Your task is specifically to make changes to non-test files in the current directory in order to fix the issue described in the PR description in a way that is general and consistent with the codebase.
-
-    <IMPORTANT>This is an interactive process where you will think and issue ONE command, see its result, then think and issue your next command.</IMPORTANT>
+    <IMPORTANT>This is an interactive process where you will think and issue AT LEAST ONE command, see the result, then think and issue your next command(s).</important>
 
     For each response:
 
@@ -54,77 +39,30 @@ agent:
 
     You are operating in an environment where
 
-    1. You write a single command
-    2. The system executes that command in a subshell
-    3. You see the result
-    4. You write your next command
+    1. You issue at least one command
+    3. The system executes the command(s) in a subshell
+    4. You see the result(s)
+    5. You write your next command(s)
 
     Each response should include:
 
-    1. A **THOUGHT** section where you explain your reasoning and plan
-    2. A single bash code block with your command
-
-    Format your responses like demonstrated within the <format_example> block:
-
-    <format_example>
-    THOUGHT: Here I explain my reasoning process, analysis of the current situation,
-    and what I'm trying to accomplish with the command below.
-
-    ```mswea_bash_command
-    your_command_here
-    ```
-    </format_example>
-
-    Commands must be specified in a single bash code block:
-
-    ```mswea_bash_command
-    your_command_here
-    ```
+    1. **Reasoning text** where you explain your analysis and plan
+    2. At least one tool call with your command
 
     **CRITICAL REQUIREMENTS:**
 
-    - Your response SHOULD include a THOUGHT section explaining your reasoning
-    - Your response MUST include EXACTLY ONE bash code block
-    - This bash block MUST contain EXACTLY ONE command (or a set of commands connected with && or ||)
-    - If you include zero or multiple bash blocks, or no command at all, YOUR RESPONSE WILL FAIL
-    - Do NOT try to run multiple independent commands in separate blocks in one response
+    - Your response SHOULD include reasoning text explaining what you're doing
+    - Your response MUST include AT LEAST ONE bash tool call
     - Directory or environment variable changes are not persistent. Every action is executed in a new subshell.
     - However, you can prefix any action with `MY_ENV_VAR=MY_VALUE cd /path/to/working/dir && ...` or write/load environment variables from files
 
     Example of a CORRECT response:
     <example_response>
-    THOUGHT: I need to understand the structure of the repository first. Let me check what files are in the current directory to get a better understanding of the codebase.
-
-    ```mswea_bash_command
-    ls -la
-    ```
-    </example_response>
-
-    Example of an INCORRECT response:
-
-    <example_response>
-    THOUGHT: I need to examine the codebase and then look at a specific file. I'll run multiple commands to do this.
+    I need to understand the structure of the repository first. Let me check what files are in the current directory to get a better understanding of the codebase.
 
-    ```mswea_bash_command
-    ls -la
-    ```
-
-    Now I'll read the file:
-
-    ```mswea_bash_command
-    cat file.txt
-    ```
+    [Makes bash tool call with {"command": "ls -la"} as arguments]
     </example_response>
 
-    If you need to run multiple commands, either:
-
-    1. Combine them in one block using && or ||
-    ```mswea_bash_command
-    command1 && command2 || echo "Error occurred"
-    ```
-
-    2. Wait for the first command to complete, see its output, then issue the next command in your following response.
-
     ## Environment Details
 
     - You have a full Linux shell environment
@@ -159,7 +97,7 @@ agent:
     Step 3: Submit (EXACT command required)
     You MUST use this EXACT command to submit:
 
-    ```mswea_bash_command
+    ```bash
     echo COMPLETE_TASK_AND_SUBMIT_FINAL_OUTPUT && cat patch.txt
     ```
 
@@ -187,45 +125,28 @@ environment:
 
 model:
   observation_template: |
-    {% if output.exception_info -%}
-    <exception>{{output.exception_info}}</exception>
-    {% endif -%}
-    <returncode>{{output.returncode}}</returncode>
-    {% if output.output | length < 10000 -%}
-    <output>
-    {{ output.output -}}
-    </output>
+    {%- if output.output | length < 10000 -%}
+    {
+      "returncode": {{ output.returncode }},
+      "output": {{ output.output | tojson }}
+      {%- if output.exception_info %}, "exception_info": {{ output.exception_info | tojson }}{% endif %}
+    }
     {%- else -%}
-    <warning>
-    The output of your last command was too long.
-    Please try a different command that produces less output.
-    If you're looking at a file you can try use head, tail or sed to view a smaller number of lines selectively.
-    If you're using grep or find and it produced too much output, you can use a more selective search pattern.
-    If you really need to see something from the full command's output, you can redirect output to a file and then search in that file.
-    </warning>
-    {%- set elided_chars = output.output | length - 10000 -%}
-    <output_head>
-    {{ output.output[:5000] }}
-    </output_head>
-    <elided_chars>
-    {{ elided_chars }} characters elided
-    </elided_chars>
-    <output_tail>
-    {{ output.output[-5000:] }}
-    </output_tail>
+    {
+      "returncode": {{ output.returncode }},
+      "output_head": {{ output.output[:5000] | tojson }},
+      "output_tail": {{ output.output[-5000:] | tojson }},
+      "elided_chars": {{ output.output | length - 10000 }},
+      "warning": "Output too long."
+      {%- if output.exception_info %}, "exception_info": {{ output.exception_info | tojson }}{% endif %}
+    }
     {%- endif -%}
   format_error_template: |
-    Please always provide EXACTLY ONE action in triple backticks, found {{actions|length}} actions.
+    Tool call error. Every response needs to use the 'bash' tool at least once to execute commands.
 
-    Please format your action in triple backticks as shown in <response_example>.
-
-    <response_example>
-    Here are some thoughts about why you want to perform the action.
-
-    ```mswea_bash_command
-    <action>
-    ```
-    </response_example>
+    Call the bash tool with your command as the argument:
+    - Tool: bash
+    - Arguments: {"command": "your_command_here"}
 
     If you have completed your assignment, please consult the first message about how to
     submit your solution (you will not be able to continue working on this task after that).
\end{lstlisting}

\end{document}